\documentclass[journal]{IEEEtran}
\usepackage{cite}
\ifCLASSINFOpdf
 \usepackage[pdftex]{graphicx}
 \graphicspath{{../pdf/}{../jpeg/}}
 \DeclareGraphicsExtensions{.pdf,.jpeg,.png}
\else
\fi
\usepackage{amsmath}
\usepackage{array}
\newcolumntype{L}[1]{>{\raggedright\let\newline\\\arraybackslash\hspace{0pt}}m{#1}}
\newcolumntype{C}[1]{>{\centering\let\newline\\\arraybackslash\hspace{0pt}}m{#1}}
\newcolumntype{R}[1]{>{\raggedleft\let\newline\\\arraybackslash\hspace{0pt}}m{#1}}
\usepackage{multirow}

\ifCLASSOPTIONcompsoc
  \usepackage[caption=false,font=normalsize,labelfont=sf,textfont=sf]{subfig}
\else
  \usepackage[caption=false,font=footnotesize]{subfig}
\fi
\usepackage{url}
\begin{document}
%
% paper title
% Titles are generally capitalized except for words such as a, an, and, as,
% at, but, by, for, in, nor, of, on, or, the, to and up, which are usually
% not capitalized unless they are the first or last word of the title.
% Linebreaks \\ can be used within to get better formatting as desired.
% Do not put math or special symbols in the title.
\title{Music Popularity: Metrics, Characteristics, and Audio-based Prediction}
%
%
% author names and IEEE memberships
% note positions of commas and nonbreaking spaces ( ~ ) LaTeX will not break
% a structure at a ~ so this keeps an author's name from being broken across
% two lines.
% use \thanks{} to gain access to the first footnote area
% a separate \thanks must be used for each paragraph as LaTeX2e's \thanks
% was not built to handle multiple paragraphs
%

\author{Junghyuk~Lee
        and~Jong-Seok~Lee,~\IEEEmembership{Senior Member,~IEEE}% <-this % stops a space
%\thanks{M. Shell was with the Department
%of Electrical and Computer Engineering, Georgia Institute of Technology, Atlanta,
%GA, 30332 USA e-mail: (see http://www.michaelshell.org/contact.html).}% <-this % stops a space
%\thanks{J. Doe and J. Doe are with Anonymous University.}% <-this % stops a space
\thanks{Manuscript received November 3, 2017; revised February 19, 2018.}
\thanks{This research was supported by the MISP (Ministry of Science, ICT and Future Planning), Korea, under the ``IT Consilience Creative Program" (IITP-2017-2017-0-01015) supervised by the IITP (Institute for Information \& Communications Technology Promotion).}
\thanks{A preliminary version of this work was presented at the Workshop on Speech, Language and Audio in Multimedia (SLAM) in 2015 \cite{slam2015}.}
\thanks{\textsuperscript{\textcopyright} 2018 IEEE. Personal use of this material is permitted. Permission from IEEE must be obtained for all other uses, in any current or future media, including reprinting/republishing this material for advertising or promotional purposes, creating new collective works, for resale or redistribution to servers or lists, or reuse of any copyrighted component of this work in other works.}
\thanks{This is the author’s version of an article that has been published in this
journal (DOI: 10.1109/TMM.2018.2820903). Changes were made to this
version by the publisher prior to publication. The final version of record is
available at \protect\url{https://doi.org/10.1109/TMM.2018.2820903}.}
}

% note the % following the last \IEEEmembership and also \thanks - 
% these prevent an unwanted space from occurring between the last author name
% and the end of the author line. i.e., if you had this:
% 
% \author{....lastname \thanks{...} \thanks{...} }
%                     ^------------^------------^----Do not want these spaces!
%
% a space would be appended to the last name and could cause every name on that
% line to be shifted left slightly. This is one of those "LaTeX things". For
% instance, "\textbf{A} \textbf{B}" will typeset as "A B" not "AB". To get
% "AB" then you have to do: "\textbf{A}\textbf{B}"
% \thanks is no different in this regard, so shield the last } of each \thanks
% that ends a line with a % and do not let a space in before the next \thanks.
% Spaces after \IEEEmembership other than the last one are OK (and needed) as
% you are supposed to have spaces between the names. For what it is worth,
% this is a minor point as most people would not even notice if the said evil
% space somehow managed to creep in.

% The paper headers
\markboth{Journal of \LaTeX\ Class Files,~Vol.~14, No.~8, August~2015}%
{Shell \MakeLowercase{\textit{et al.}}: Bare Demo of IEEEtran.cls for IEEE Journals}
% The only time the second header will appear is for the odd numbered pages
% after the title page when using the twoside option.
% 
% *** Note that you probably will NOT want to include the author's ***
% *** name in the headers of peer review papers.                   ***
% You can use \ifCLASSOPTIONpeerreview for conditional compilation here if
% you desire.

% If you want to put a publisher's ID mark on the page you can do it like
% this:
%\IEEEpubid{0000--0000/00\$00.00~\copyright~2015 IEEE}
% Remember, if you use this you must call \IEEEpubidadjcol in the second
% column for its text to clear the IEEEpubid mark.

% use for special paper notices
%\IEEEspecialpapernotice{(Invited Paper)}

% make the title area
\maketitle

% As a general rule, do not put math, special symbols or citations
% in the abstract or keywords.
\begin{abstract}
Understanding music popularity is important not only for the artists who create and perform music but also for music-related industry. 
It has not been studied well how music popularity can be defined, what are its characteristics, and whether it can be predicted, which are addressed in this paper. %is ambiguous and not well defined about how music popularity is and what characteristic of the popularity is.
We first define eight popularity metrics to cover multiple aspects of popularity. 
Then, analysis of each popularity metric is conducted with long-term real-world chart data to deeply understand the characteristics of music popularity in the real world. 
We also build classification models for predicting popularity metrics using acoustic data. %from audio signal processing then discuss the results.
In particular, we focus on evaluating features describing music complexity together with other conventional acoustic features including MPEG-7 and Mel-frequency cepstral coefficient (MFCC) features. 
Results show that, although there exists still room for improvement, it is feasible to predict the popularity metrics of a song significantly better than random chance based on its audio signal, particularly using both the complexity and MFCC features.
%there is good opportunity to predict popularity of a song based on its audio signal.
%
%According to the prior works about relation between musical complexity and preference, complexity of a song is proposed as acoustic classification features in our work as well as MFCC and MPEG-7 features which are well established acoustic feature extraction methods.
%We compare the prediction performance of each feature, combine the features to boost the performance, and examine additional performance enhancing method.
%Different features showing good performance for different popularity patterns, it reaches highest when the Complexity feature is added to the MFCC feature.

\end{abstract}

% Note that keywords are not normally used for peerreview papers.

% For peer review papers, you can put extra information on the cover
% page as needed:
% \ifCLASSOPTIONpeerreview
% \begin{center} \bfseries EDICS Category: 3-BBND \end{center}
% \fi
%
% For peerreview papers, this IEEEtran command inserts a page break and
% creates the second title. It will be ignored for other modes.
\IEEEpeerreviewmaketitle

\section{Introduction} \label{Intro}
% The very first letter is a 2 line initial drop letter followed
% by the rest of the first word in caps.
% 
% form to use if the first word consists of a single letter:
% \IEEEPARstart{A}{demo} file is ....
% 
% form to use if you need the single drop letter followed by
% normal text (unknown if ever used by the IEEE):
% \IEEEPARstart{A}{}demo file is ....
% 
% Some journals put the first two words in caps:
% \IEEEPARstart{T}{his demo} file is ....
% 
% Here we have the typical use of a "T" for an initial drop letter
% and "HIS" in caps to complete the first word.
%\IEEEPARstart{T}{his} demo file is intended to serve as a ``starter file''
%for IEEE journal papers produced under \LaTeX\ using
%IEEEtran.cls version 1.8b and later.
%% You must have at least 2 lines in the paragraph with the drop letter
%% (should never be an issue)
%I wish you the best of success.
%
%\hfill mds
% 
%\hfill August 26, 2015

%\subsection{Subsection Heading Here}
%Subsection text here.

%% needed in second column of first page if using \IEEEpubid
%%\IEEEpubidadjcol
%
%\subsubsection{Subsubsection Heading Here}
%Subsubsection text here.

\IEEEPARstart{I}{n} these days, a large amount of multimedia content is delivered to users through various media and platforms.
The popularity of multimedia content items has been considered important, because it plays a critical role to deal with various issues of content management such as recommendation, search and retrieval, network content caching, advertisement, etc. 
Thus, there have been significant research efforts to understand and predict content popularity, especially for images and videos \cite{vidpop2013, vidpop2017, viddyn2015, viddyn2016, impop2013, friend2016, caching2015}.

Music is a kind of content that has similar issues and challenges. 
Digital music sales have increased significantly. 
Paid subscriptions and industry revenues from streaming are more than tripled between 2011 and 2014 \cite{sales2014}. 
In this large market, some songs are popular and some others are not. 
Popularity of a song can be measured as its total sales or exposure to the public, and is often summarized in a music chart. 
For instance, the Billboard charts determine the ranking of songs based on total sales, online streaming counts, etc. on a weekly basis; last.fm provides information about top tracks and artists from live radio plays. 
Using a music chart, it is possible to understand not only the current ranking of a song but also long-term popularity changes such as whether the song has been steadily popular over a period. 

The companies and individuals involved in the music production, distribution, and management industry are very interested in music popularity. 
Sometimes they want to know popularity of a song in advance of its release to the public. 
Prediction of popular songs can be used in audio source management for online streaming services and music delivery services. 
For example, it would be effective to generate a playlist \cite{playlist2001, playlist2002} or recommend a song \cite{recomm2004, recomm2005, recomm2008, recomm2014} predicted to become popular to a user who usually enjoys popular songs. 
As another example, it would be useful to use a song predicted to become popular in advertisement in order to enhance the marketing effect due to the public response to the song. 

There exist a few studies investigating the evolution of popular music. 
In \cite{evolution2012}, based on simulation with a Darwinian model, it was shown that competing evolutionary forces can explain the dynamics of public music preference. 
In \cite{evolution2012west}, evolving trends of western popular music were revealed, which appeared in pitch, timbre, and loudness. 
Similarly, in \cite{evolution2015}, it was shown that popular music has evolved continuously and even sometimes abruptly in terms of genre distribution, diversity, and timbre.

In an early psychological study, it has been proposed that the musical complexity affects listner's preference on the song \cite{u1975}.
In \cite{complexity2004}, it was examined how much musical features describing complexity and chart performance are correlated. 
The rhythmic complexity and melodic complexity of songs were measured using temporal interval and pitch changes of notes, respectively. 
Four metrics of chart performance such as the total number of weeks, average rank change, peak ranking, and debut ranking were considered. 
For the Billboard Modern Rock Top 40 chart for six months, it was concluded the musical features and chart performance have moderate correlations. 
However, this study is very limited in that it considered only ten songs with their popularity data only for a half year, and used MIDI files of the songs instead of the original audio signals. 
Thus, its conclusions may not be reliable. 
Furthermore, the study did not suggest any popularity prediction models. 
As an attempt to predict popularity of music, in \cite{hit2005}, a classification model was developed, which predicted using acoustic features and lyric features whether a song would be ranked highest in a music chart. 
Mel-frequency cepstral coefficients (MFCCs) were used as acoustic features, and the semantic information was extracted from lyrics. 
Support vector machine (SVM)-based classification for the songs on the Oz Net Music Chart Trivia Page was conducted, which showed that music popularity was not easily predictable using acoustic features, although lyric-based features were slightly effective. 
Similarly, in \cite{hit2012}, it was shown that a variety of acoustic features including MPEG-7 Audio Standard features were not effective to classify songs into three popularity categories (high, medium, and low). 
Recently, there were attempts to apply deep learning to predict the hit score of a song \cite{yang2017CNN, yu2017CNN}.
They used convolutional neural networks (CNNs) with the Mel-spectrogram of a song as an input feature map to predict the playcount of the song in a music streaming service. 
It was shown that the CNNs are more effective than shallow models.

Some studies considered information other than acoustic features. 
In fact, social factors sometimes play an important role in determining whether a song would be popular or not \cite{market2006}.
In \cite{season2010}, it was shown that seasonal music preference has significant impact on music popularity (e.g., Christmas carols in December).
In \cite{twitter2013}, the social media information, i.e., the number of Twitter posts with hashtags indicating that users were listening to particular songs, was considered. 
Prediction of weekly rankings of songs for 10 weeks in the Billboard Hot 100 chart was conducted as classification into decadal rank groups (i.e., 1 to 10, 11 to 20, etc.), which was more successful than the case with only acoustic features. 
However, this approach has a limitation in the perspective of popularity prediction because public responses to a song are available only after it is distributed. 

In comparison to the aforementioned studies, this paper aims at providing more comprehensive understanding and deeper insight into predicting music popularity based on acoustic information using a large-scale data set. 
The research questions considered are:\begin{itemize}
\item How can the popularity of a song over time be described from a music chart?
\item What are recognizable characteristics of popularity metrics in long-term real-world chart data?
\item To what extent can the popularity metrics of a song be predicted from the audio signal?
\item What are effective audio features showing good prediction performance?
\end{itemize}

Our distinguished contributions to answer these questions can be summarized as follows: \begin{enumerate}
\item We define popularity metrics that summarize music popularity in various viewpoints. 
As explained above, most of the previous studies considered only one aspect of popularity from charts. 
However, the chart performance of a song is in fact a temporal sequence, and thus summarizing it by one single value ignores other aspects of popularity. 
Therefore, we define eight popularity metrics that reflect diverse popularity aspects. 
\item We provide in-depth analysis of the popularity metrics for large-scale real chart data in order to identify significant characteristics of real-world music popularity. 
While many studies have investigated popularity patterns for video (e.g., \cite{viddyn2015, viddyn2016, vidchar2013}), there exist few studies for music popularity. 
We examine statistics of each metric, relationship between metrics, and temporal evolution of the metrics.
\item Various acoustic features are benchmarked for prediction of the popularity metrics. 
In particular, we focus on validating music complexity features for popularity metric prediction on the ground that musical complexity affects musical preference and also popularity. 
\end{enumerate}

The rest of this paper is organized as follows.
In Section~\ref{Popularity Metrics}, we propose the definitions of the proposed popularity metrics. 
Next, in Section~\ref{Analysis}, we analyze the popularity metrics appearing in the Billboard Hot 100 chart for the past 45 years. 
In Section~\ref{Prediction}, we present our experimental results of audio signal-based popularity prediction approaches. 
Finally, Section~\ref{Conclusions} provides concluding remarks. 

\section{Popularity metrics} \label{Popularity Metrics}

Popularity of a song can be described in various perspectives such as how many times the song has been sold, how many times the song has been broadcasted on TV, how many times the song has been queried in a search engine, how many times the song has been consumed via streaming, etc. 
A music popularity chart such as the Billboard Hot 100 chart tries to summarize such statistics on, e.g., a weekly basis, resulting in a popularity ranking of songs. 
Therefore, the ranking of a song in a chart is basically a time domain signal, representing its popularity over time. 
As mentioned in the introduction, the previous studies mostly considered only one aspect of this signal to summarize the popularity of a song, which loses information regarding the dynamic nature of the popularity. 
For instance, when the highest rank of a song on the chart is considered, it is not possible to properly identify a song that is a long-term steady-seller if its highest rank is not so high.

In this study, we define multiple popularity metrics extracted from rankings of songs over time in a music chart so that we can examine both instantaneous and dynamic aspects of popularity. 
These popularity metrics are measured using \textit{rank score}, which is the inverted rank on a chart, i.e., the rank score of song $i$ is obtained as:
\begin{equation}Rank\_score(i)=Max\_rank - rank(i) + 1\end{equation}
where $Max\_rank$ is the lowest rank of the chart and $rank(i)$ is the rank of the song. 
For instance, in a weekly top 100 chart, the song ranked highest has a rank score of 100 and the song ranked lowest has a rank score of 1.

\begin{figure*}[!t]
\centering
\subfloat[\textit{Debut}]{\includegraphics[width=1.5in]{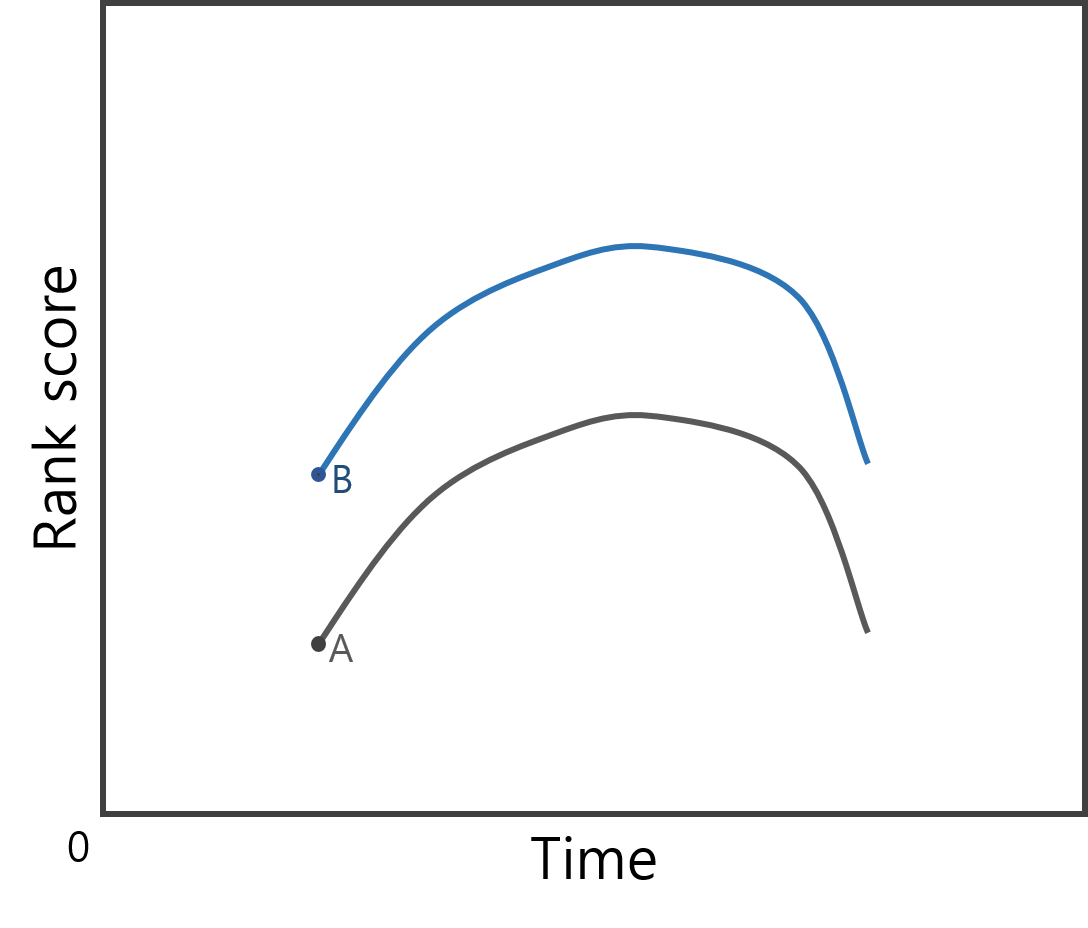}
\label{fig:debut}}
\hfil
\subfloat[\textit{Max}]{\includegraphics[width=1.5in]{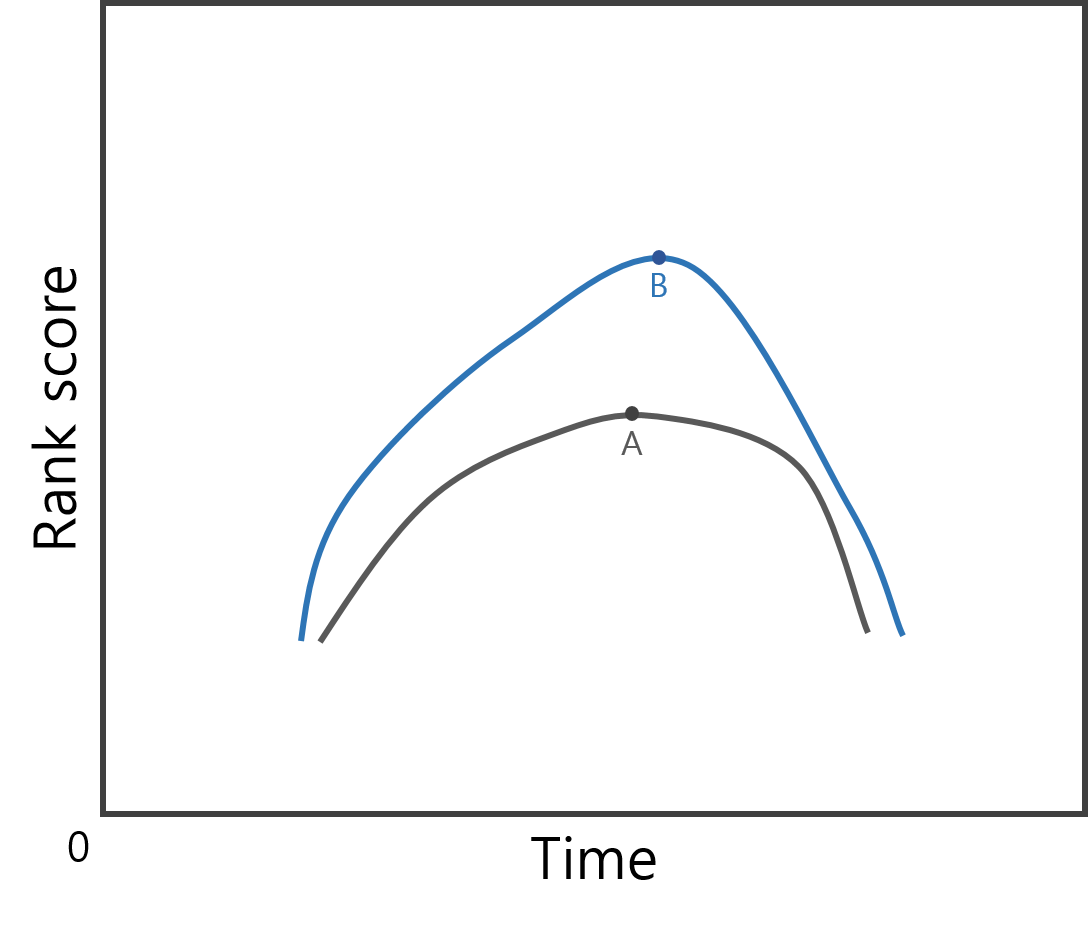}
\label{fig:max}}
\hfil
\subfloat[\textit{Mean}]{\includegraphics[width=1.5in]{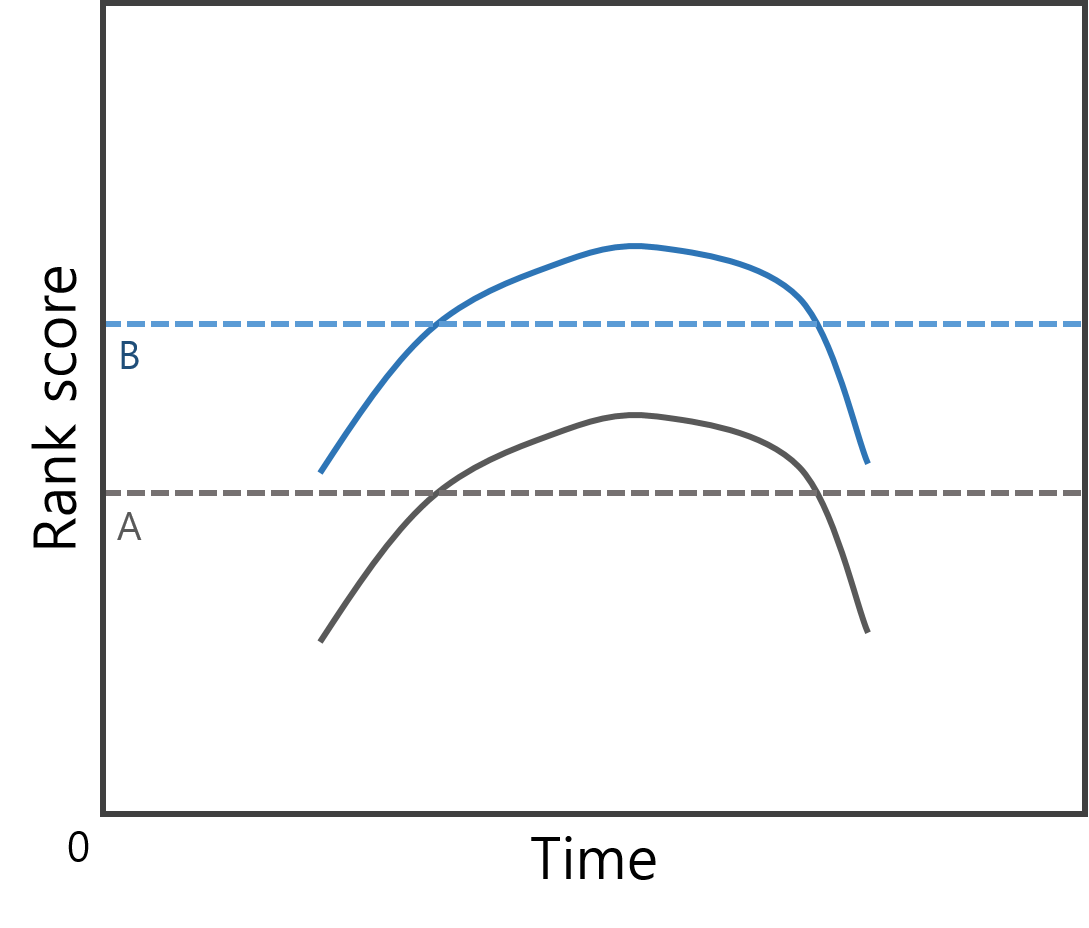}
\label{fig:mean}}
\hfil
\subfloat[\textit{Std}]{\includegraphics[width=1.5in]{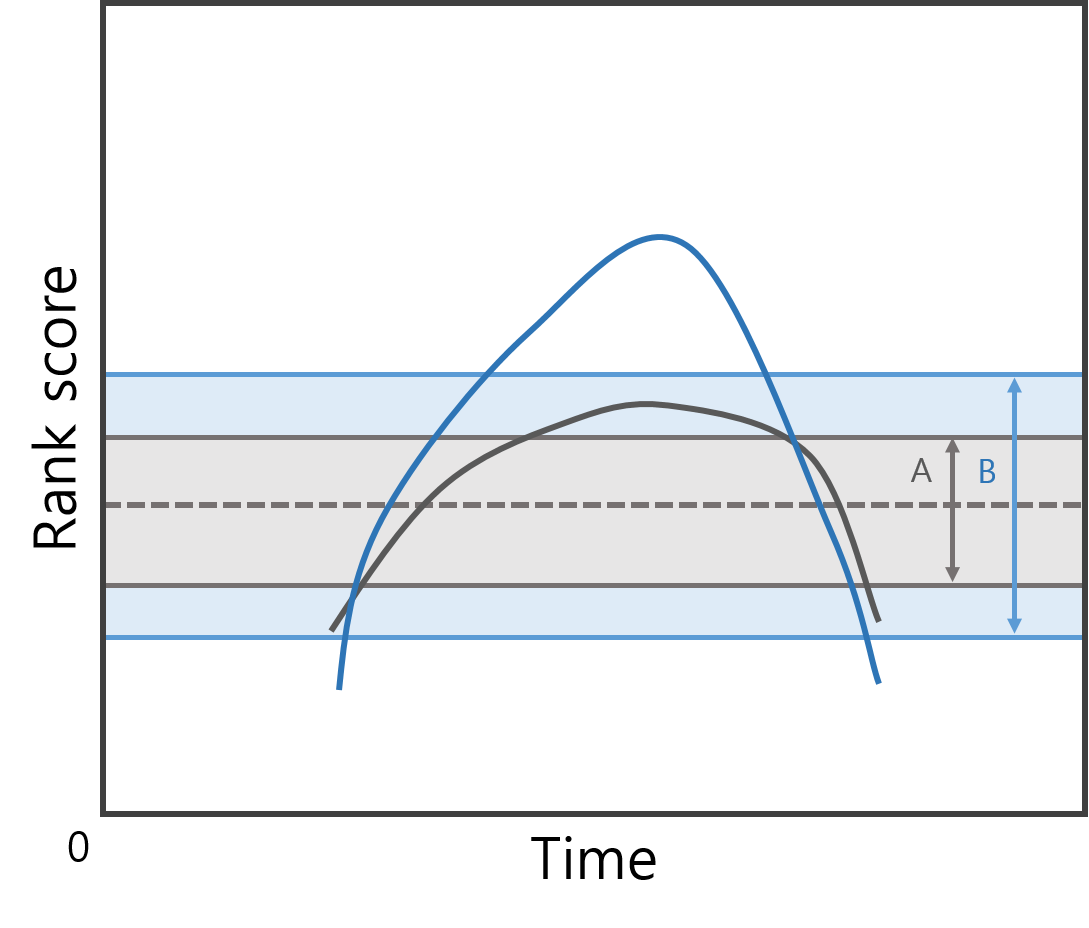}
\label{fig:std}}

\subfloat[\textit{Length}]{\includegraphics[width=1.5in]{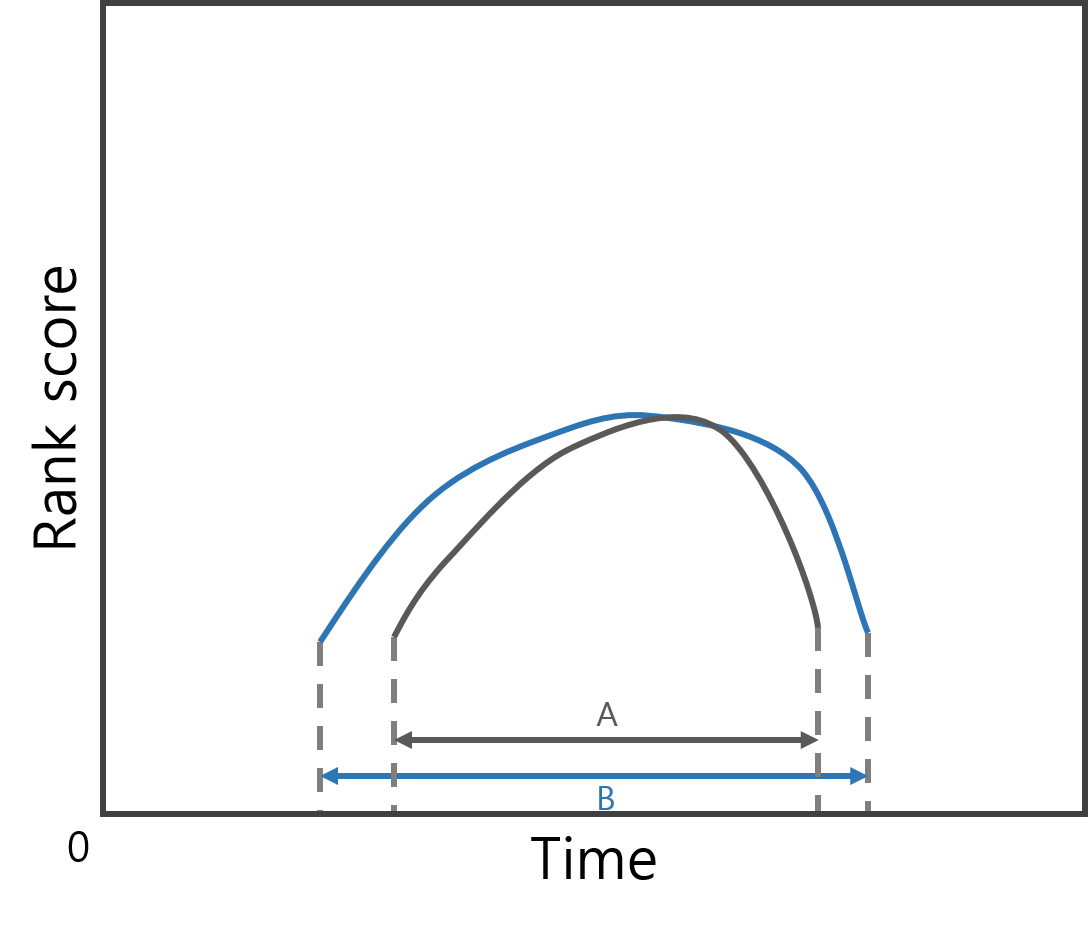}
\label{fig:length}}
\hfil
\subfloat[\textit{Sum}]{\includegraphics[width=1.5in]{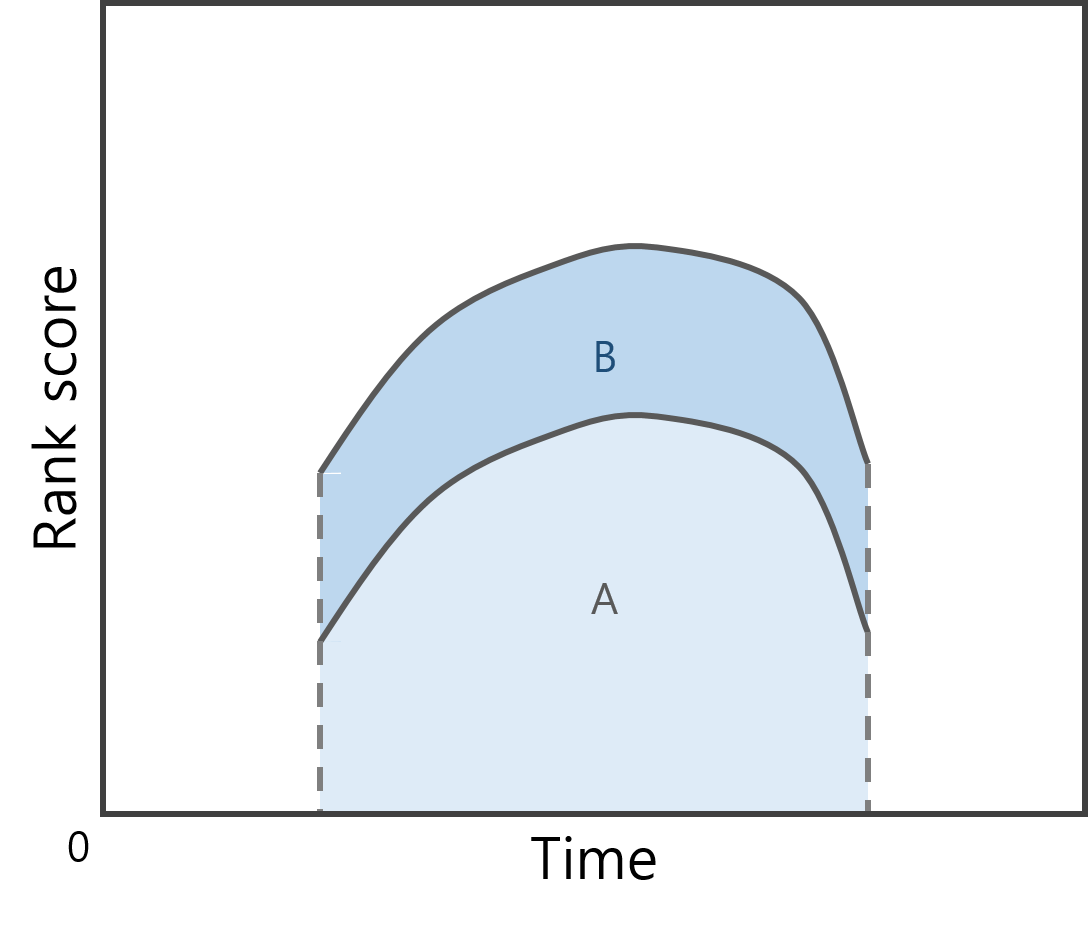}
\label{fig:sum}}
\hfil
\subfloat[\textit{Skewness}]{\includegraphics[width=1.5in]{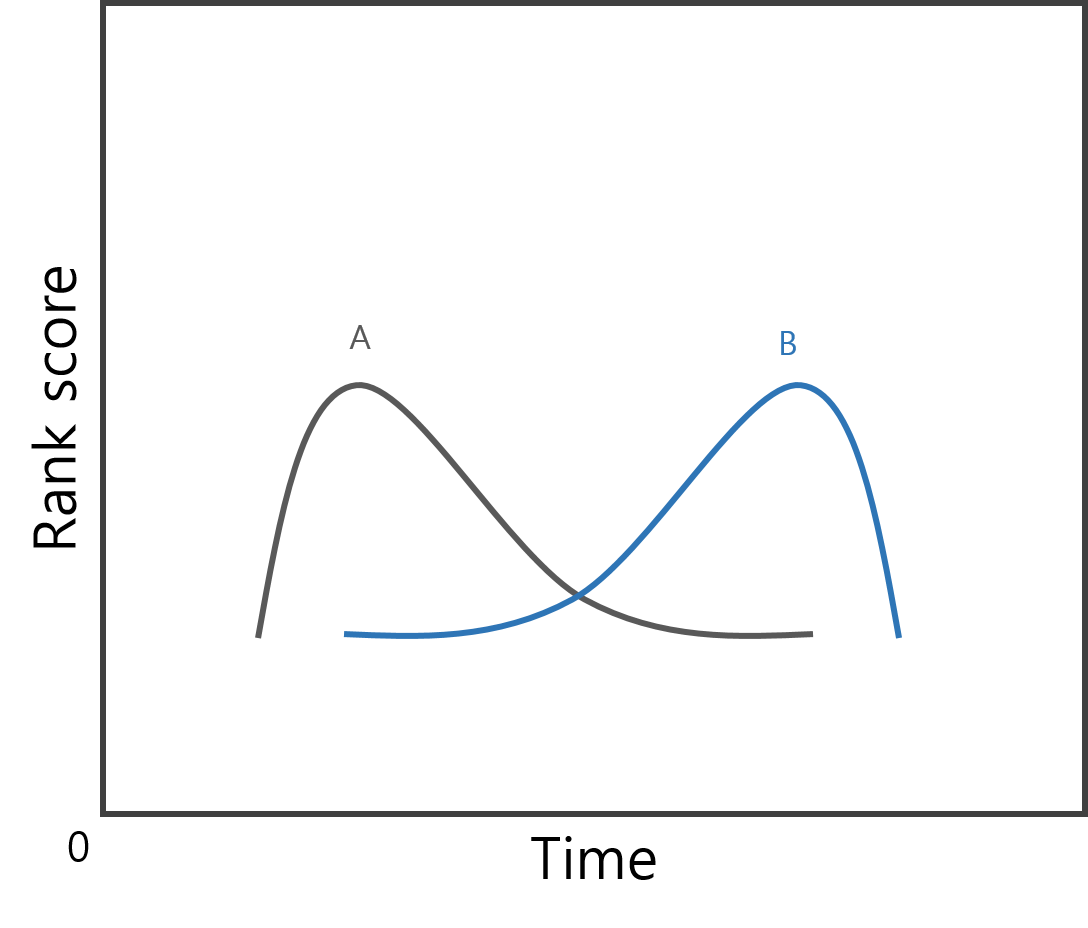}
\label{fig:skewness}}
\hfil
\subfloat[\textit{Kurtosis}]{\includegraphics[width=1.5in]{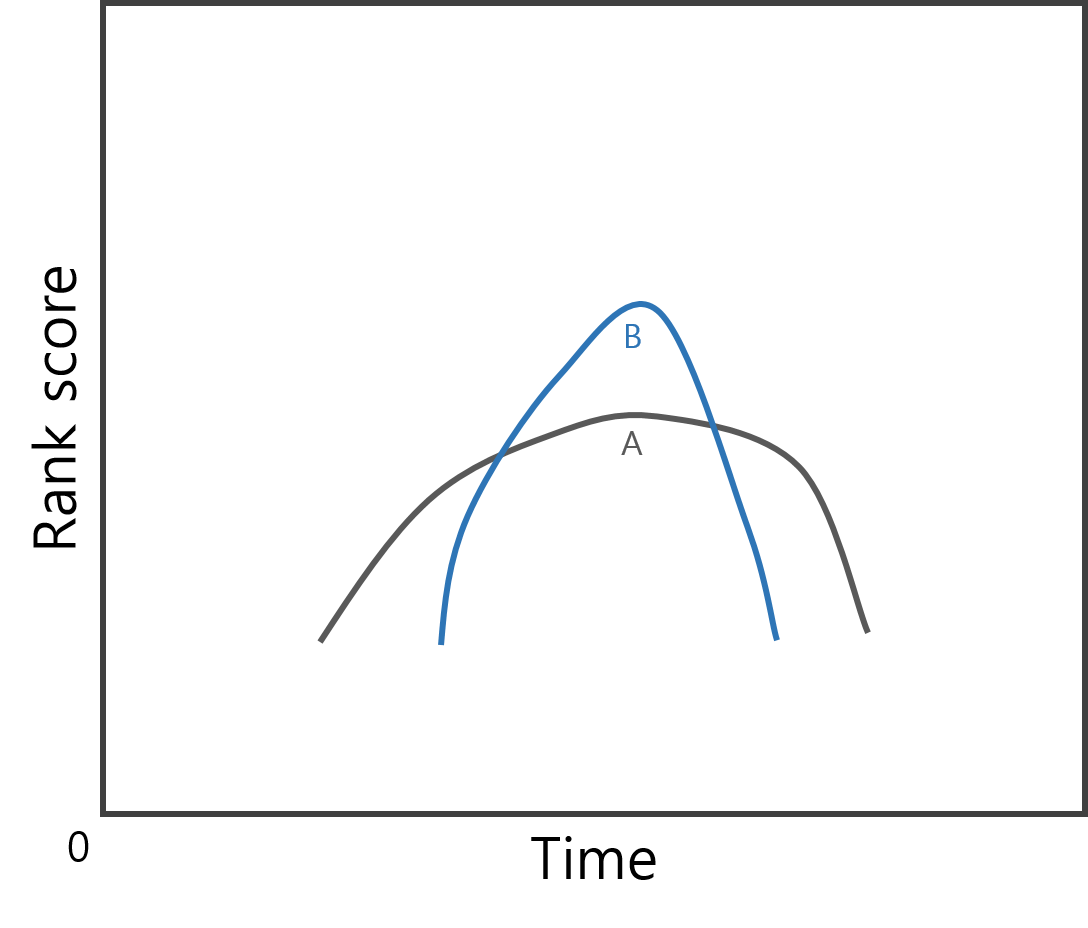}
\label{fig:kurtosis}}
\caption{Illustrations of the popularity metrics defined in our work.}
\label{fig:pps}
\end{figure*}

We define eight popularity metrics as follows (see \figurename~\ref{fig:pps}):
\begin{enumerate}[\IEEEsetlabelwidth{$\alpha\omega\pi\theta$}]
\item[\textit{Debut}] 
This is defined as the rank score of a song when the song appears first in a chart.
It indicates the initial popularity of the song.
In \figurename~\ref{fig:debut}, song B has a larger value of \textit{Debut} than song A.
\item[\textit{Max}] 
This is defined as the maximum rank score of a song during the whole period, measuring the maximum popularity of the song.
The maximum rank score of song B is higher than that of song A in \figurename~\ref{fig:max} and thus song B has a larger value of \textit{Max} than song A.
\item[\textit{Mean}] 
This is defined as the average rank score of a song over the whole period during which the song appears in a chart. 
In \figurename~\ref{fig:mean}, song B has a larger value of \textit{Mean} than song A (marked with dotted lines). 
\item[\textit{Std}] 
This is the standard deviation of the rank scores of a song over the whole period during which the song appears in a chart. 
It describes how much the popularity of a song has changed over time. 
In \figurename~\ref{fig:std}, the rank score of song B has changed more, so its value of \textit{Std }is larger than that of song A.
\item[\textit{Length}] 
This is defined as the time period (e.g., the number of weeks) during which a song appears on a chart.
It measures how long a song has been popular, and thus can identify steadily popular songs.
In \figurename~\ref{fig:length}, song B has a larger value of \textit{Length }than song A.
\item[\textit{Sum}] 
This is defined as the sum of the rank scores over time.
Although this is similar to \textit{Length}, \textit{Length} measures only the time period, whereas, \textit{Sum} considers the rank score during which a song stays in a chart.
Thus, it basically describes the overall popularity of a song when the whole time period is considered.
In \figurename~\ref{fig:sum}, song B always has higher scores, so the value of \textit{Sum }of song B is larger than that of song A.
Note that \textit{Sum} is different from \textit{Mean} because the period during which a song appears in a chart is different for different songs.
\item[\textit{Skewness}]
This is the skewness (i.e., the third moment) of the rank scores of a song. 
This partially describes the dynamic patterns that a song gains and loses popularity. 
A positive value of \textit{Skewness} (song A in \figurename~\ref{fig:skewness}) means that the song has become popular fast, reached its highest popularity, and become unpopular slowly; a negative value of \textit{Skewness} (song B in \figurename~\ref{fig:skewness}) indicates that the song has become popular slowly, then lost its popularity fast.
\item[\textit{Kurtosis}] 
Together with \textit{Skewness}, the kurtosis of a song describes the patterns of growing and declining popularity. 
The faster the popularity growth of a song is, the larger the value of \textit{Kurtosis} is. 
The rank score curve for song B in \figurename~\ref{fig:kurtosis} is sharper than that of song A, so \textit{Kurtosis }of song B is larger.
\end{enumerate}

\section{Analysis of popularity metrics} \label{Analysis}

\subsection{Data}
The Billboard magazine distributes weekly lists of popular songs, called the Billboard charts, since 1940. 
The Billboard Hot 100 chart\footnote{\url{http://www.billboard.com/charts/hot-100}} is the most representative among the charts, which provides 100 most popular songs (regardless of their genres) every week. 
The ranking in this chart is based on radio airplay, sales data, and streaming activity data from Nielsen Music\footnote{\url{http://www.nielsen.com/us/en/solutions/measurement/music-sales-measurement.html}}.

We use the chart data between 1970 and 2014, comprising 209,000 chart ranks for 2,090 weeks in total. 
There are 18,604 distinct songs in these data.
We exclude the songs that appeared only one or two weeks because some popularity metrics cannot be obtained reliably for them. 
The number of songs remained only one and two weeks are 1,077 and 841, respectively. 
Finally, we use the chart data of the remaining 16,686 songs.

\subsection{Distribution analysis}
\begin{figure*}[!t]
\centering
\subfloat[]{\includegraphics[width=2.1in]{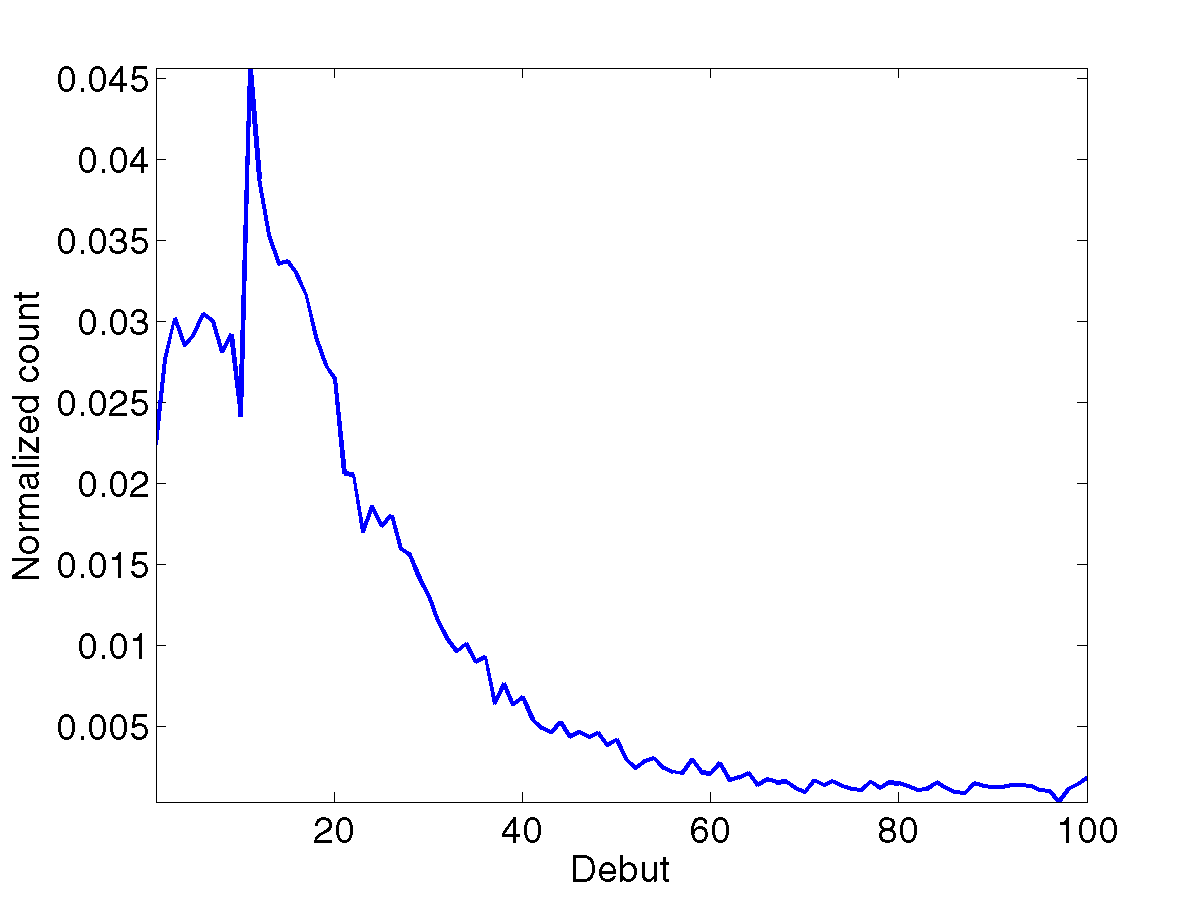}
\label{his:debut}}
\hfil
\subfloat[]{\includegraphics[width=2.1in]{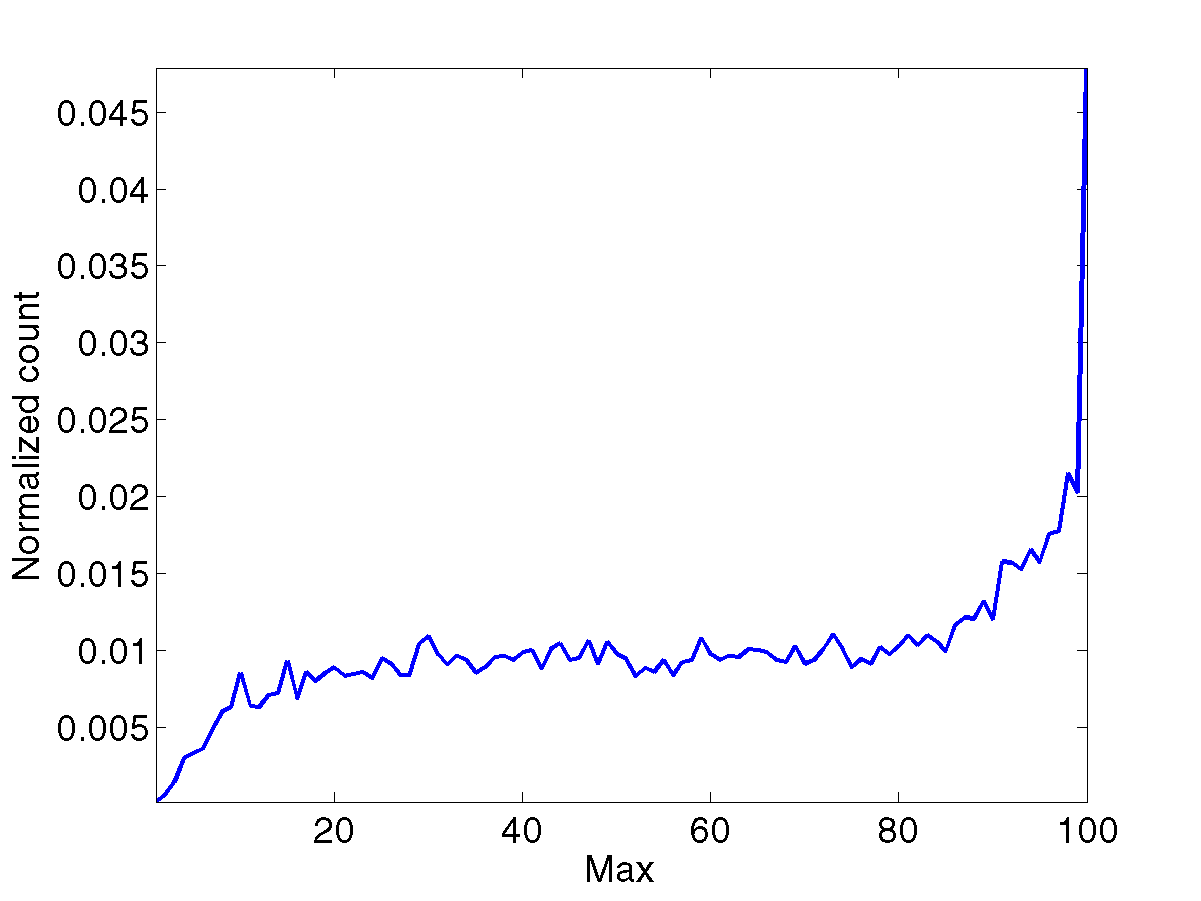}
\label{his:max}}

\subfloat[]{\includegraphics[width=2.1in]{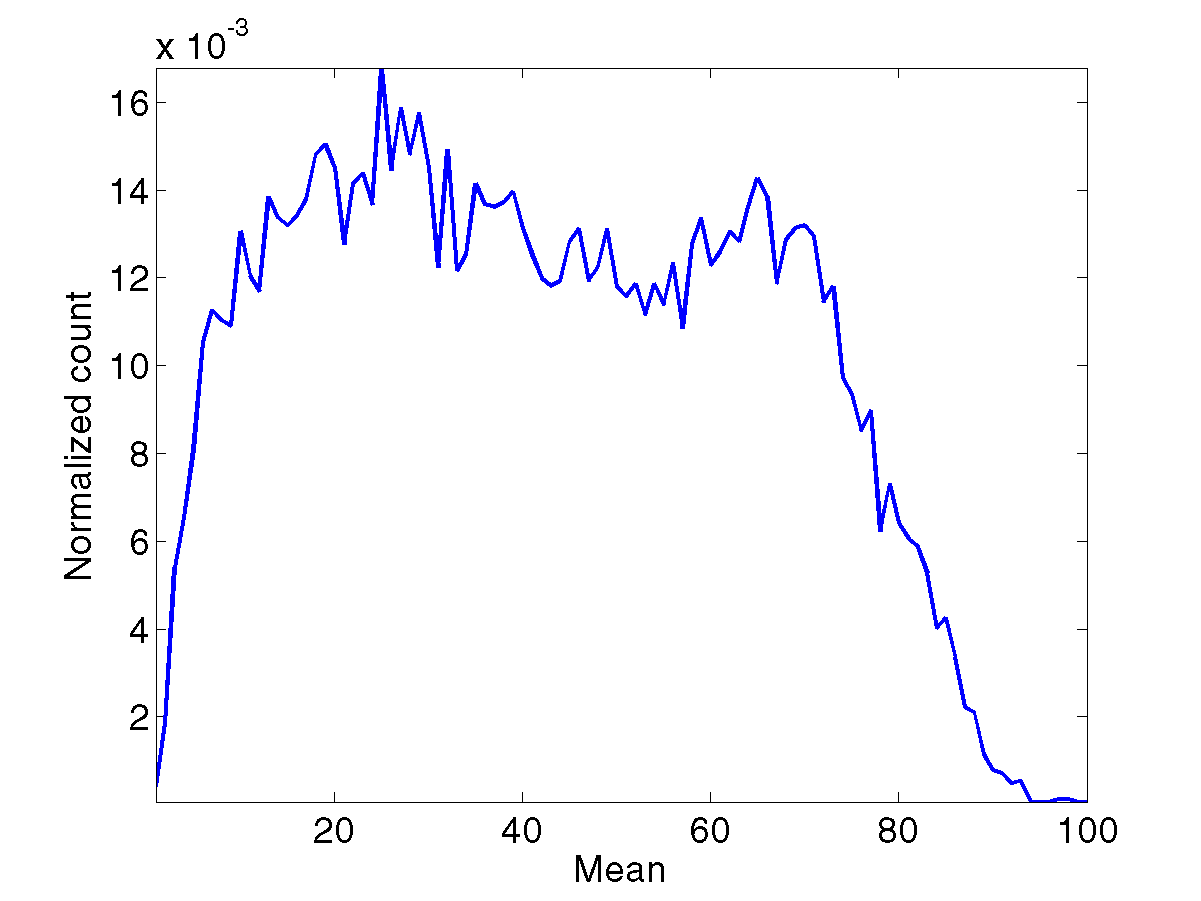}
\label{his:mean}}
\hfil
\subfloat[]{\includegraphics[width=2.1in]{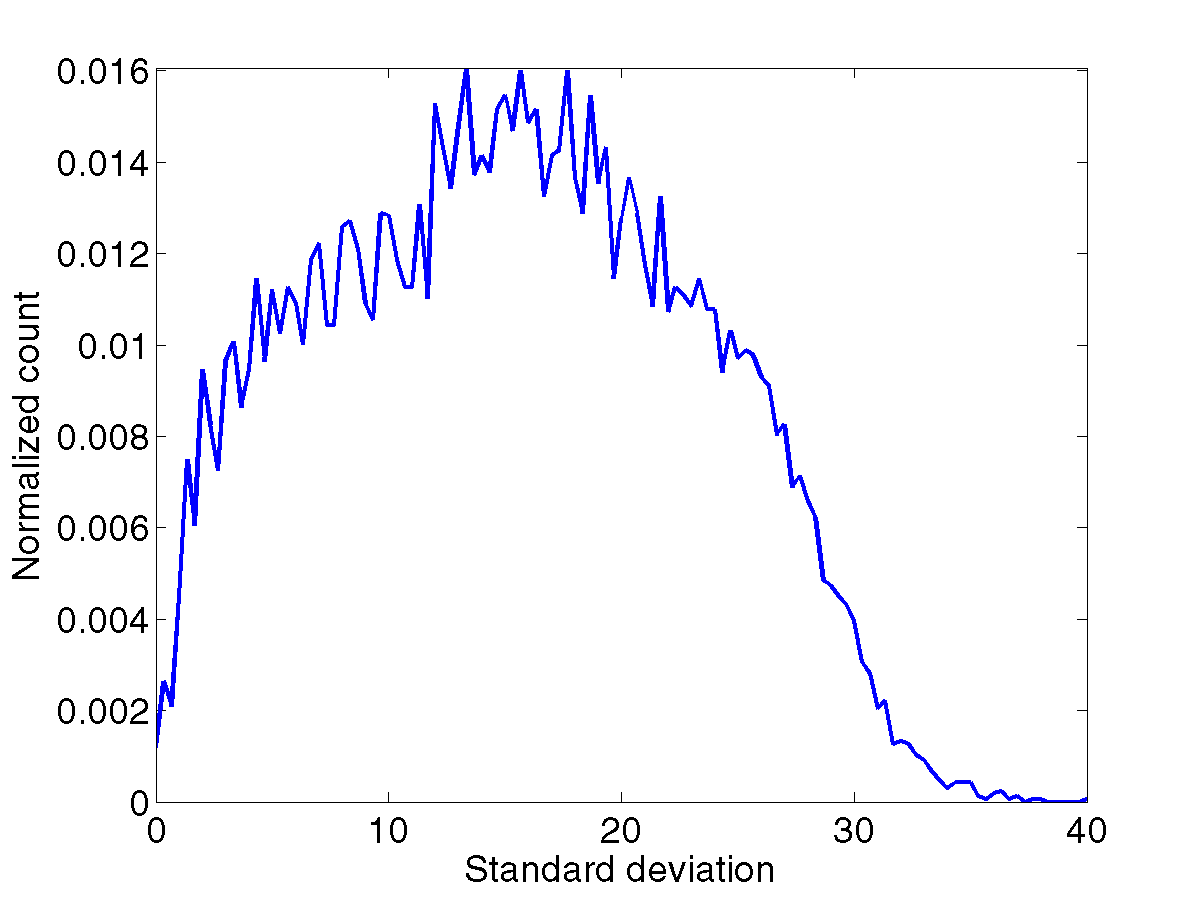}
\label{his:std}}

\subfloat[]{\includegraphics[width=2.1in]{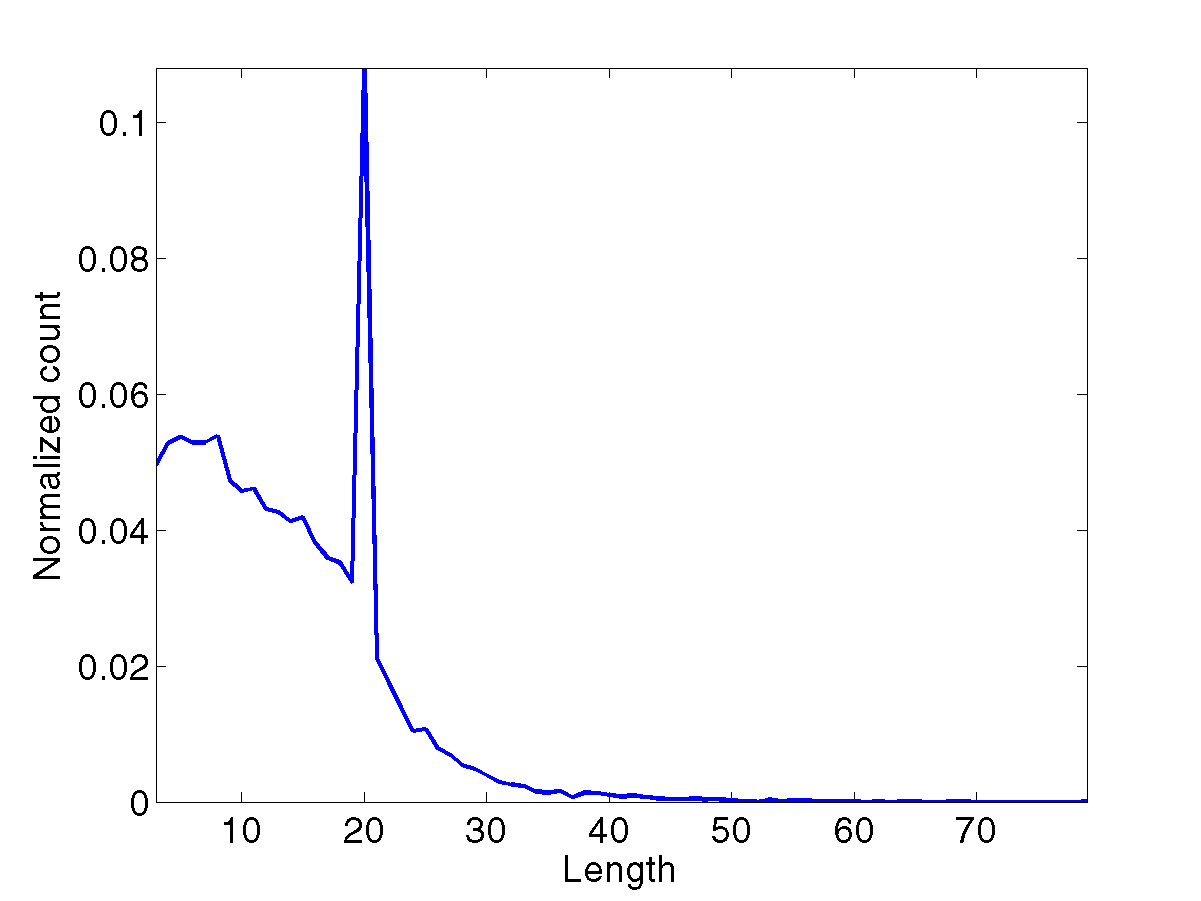}
\label{his:length}}
\hfil
\subfloat[]{\includegraphics[width=2.1in]{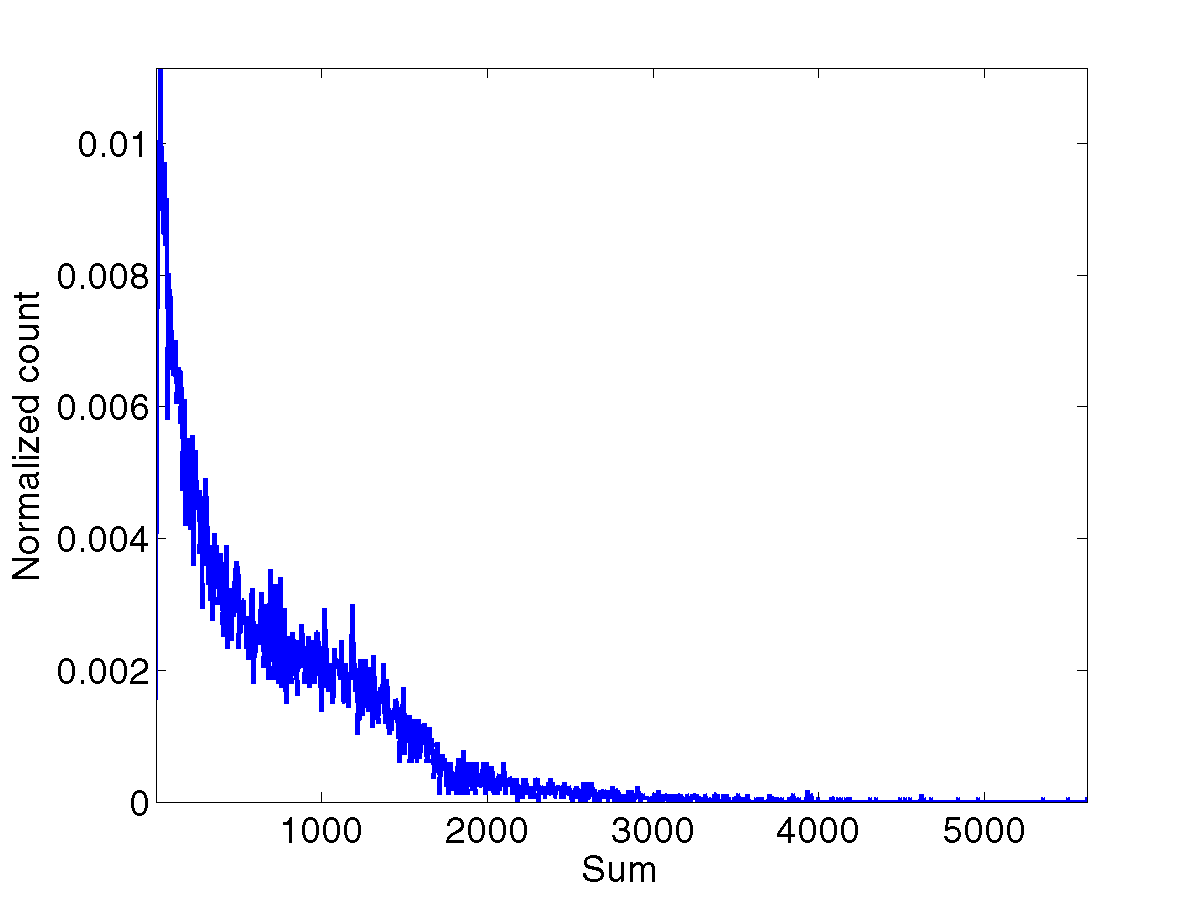}
\label{his:sum}}

\subfloat[]{\includegraphics[width=2.1in]{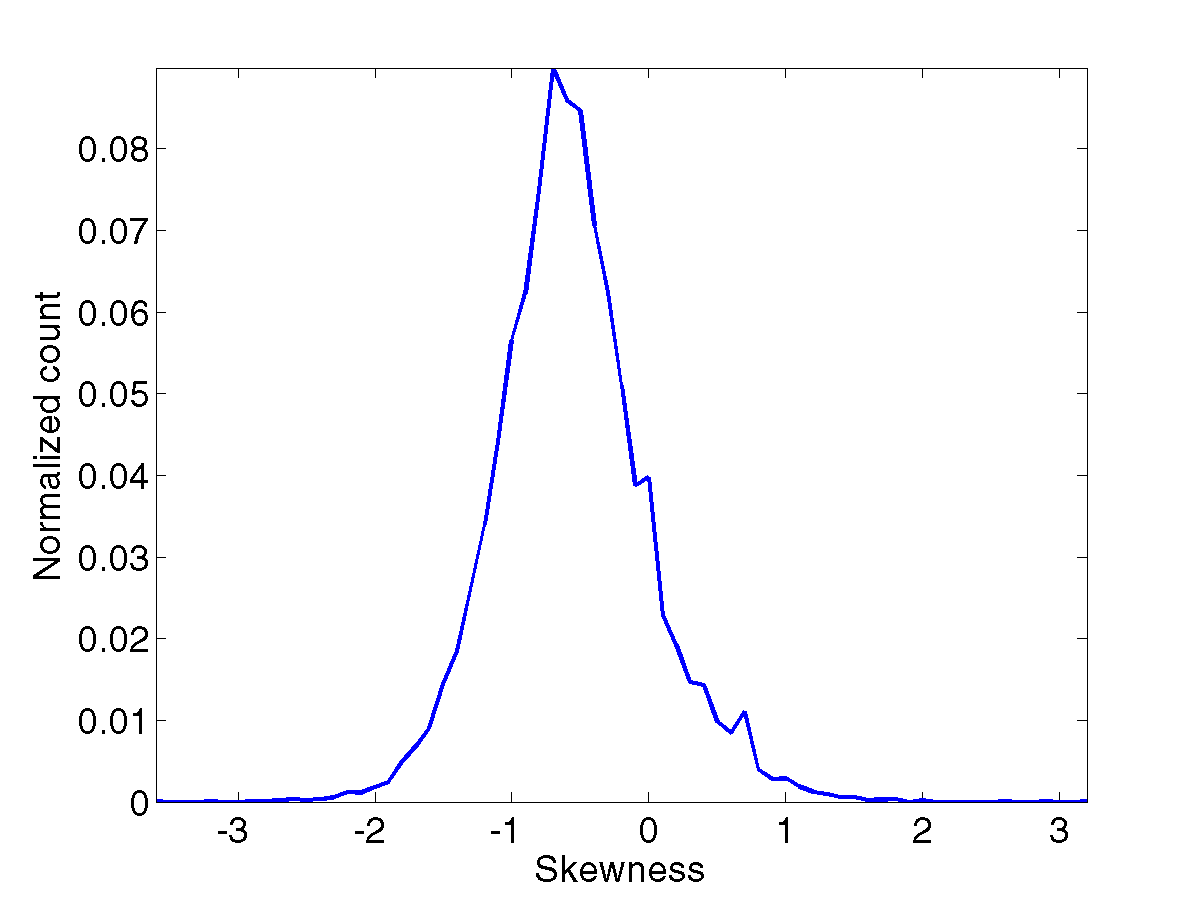}
\label{his:skewness}}
\hfil
\subfloat[]{\includegraphics[width=2.1in]{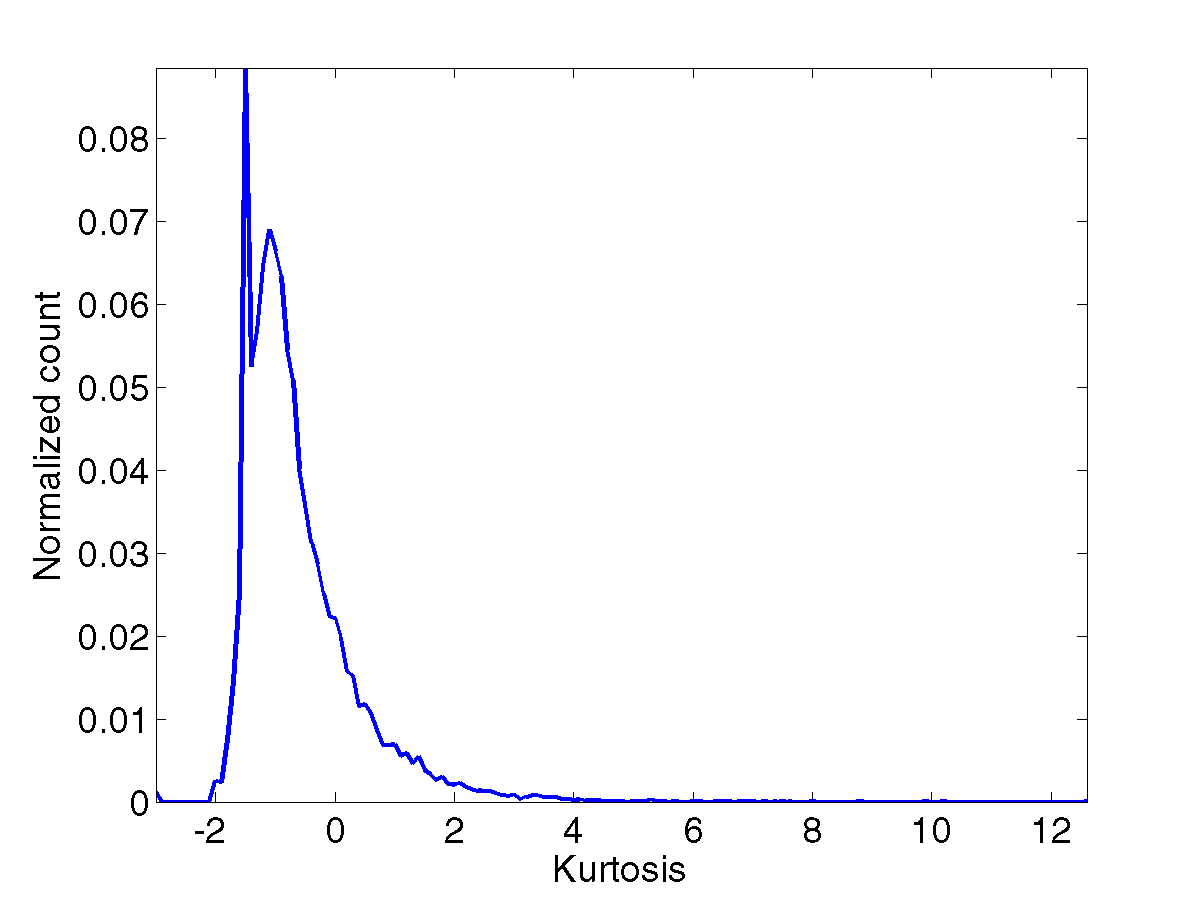}
\label{his:kurtosis}}
\caption{Histograms of popularity metrics: (a) \textit{Debut}, (b) \textit{Max}, (c) \textit{Mean}, (d) \textit{Std}, (e) \textit{Length}, (f) \textit{Sum}, (g) \textit{Skewness}, and \textit{(h) Kurtosis}.}
\label{fig:histograms}
\end{figure*}

\figurename~\ref{fig:histograms} shows the histograms of the eight popularity metrics for the whole chart data set. 

The histogram of \textit{Debut} in \figurename~\ref{his:debut} shows that most of the songs were ranked low when they appeared in the chart for the first time. 
The mean and median values of \textit{Debut} are 21.8 and 16, respectively. 
A peak at a rank score of 11 (i.e., 90th in the chart) is also observed in the histogram. 
Only a small number of songs were ranked high at the beginning; 8.1\% of the songs received \textit{Debut} scores larger than 50 (i.e., ranks higher than 51).
There also exist songs debuted as their highest ranking (less than 0.2\%).

However, in \figurename~\ref{his:max}, the histogram of \textit{Max} is concentrated in the region with high values; a peak at 100 (i.e., the top rank in the chart) is observed, showing that more than 4.5\% of songs reached the top rank.
Then, the histogram is almost flat over a wide range between about 20 and 80.

The histogram of \textit{Mean} is roughly flat in the mid-range (\figurename~\ref{his:mean}).
The mean and median values are 42.1 and 41, respectively, meaning that the average ranks of the songs are slightly lower than the middle in the chart. 
In \figurename~\ref{his:std}, \textit{Std} is distributed over a large range, with a mean of 15.3. Its maximum value is 39.9, which is relatively large.

\figurename~\ref{his:length} shows that the histogram of \textit{Length} is concentrated on small values. 
On average, the songs remained in the chart for 13.4 weeks. 
A sharp peak at 20 weeks is observed. 
This is due to the recurrent rule\footnote{\url{http://www.billboard.com/biz/billboard-charts-legend}}, which removes songs ranked below a half (50 in our case) for more than 20 weeks from the chart.
The maximum value is 79 weeks, which corresponds to ``Sail" by AWOLNATION, 2011. 

\begin{figure}[!t]
\centering
\includegraphics[scale=0.33]{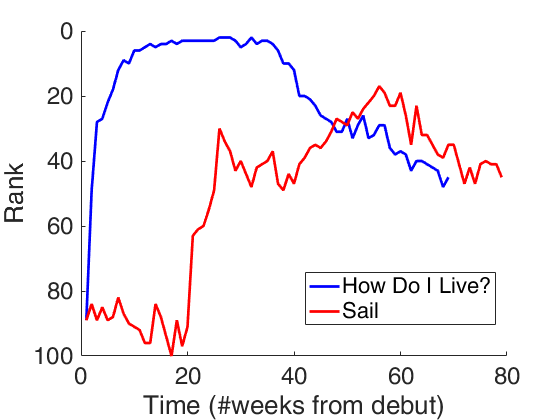}
\caption{Ranks of two songs: "How Do I Live?" (blue) and "Sail" (red).}
\label{fig:rankplot}
\end{figure}

The histogram in \figurename~\ref{his:sum} shows that many songs have small values of \textit{Sum}, and a few songs have extremely large values. 
As a result, the median and average values differ largely: 506 and 682.2, respectively.
The maximum \textit{Sum} value is 5,615 (``How Do I Live?'' by LeAnn Rimes, 1997). 
Thus, the song with the maximum value of \textit{Sum} and that with the maximum value of \textit{Length} are different.
The ranks of these two songs over time are shown in \figurename~\ref{fig:rankplot}.
The song ``Sail" stayed longer in the chart but was ranked lower than the song ``How do I live?" overall, which resulted in a lower value of \textit{Sum} of the former than the latter. 
This justifies the necessity of observing both \textit{Sum} and \textit{Length} as popularity metrics.

The histogram of \textit{Skewness} in \figurename~\ref{his:skewness} is similar to the probability density function of a Laplace distribution, with a peak at -0.7. This shows that on average, the popularity of a song typically grows slowly and falls fast. 

The histogram of \textit{Kurtosis} is also peaked at a negative value, as shown in \figurename~\ref{his:kurtosis}. 
The mean value of \textit{Kurtosis} is -0.62, and most of the songs have negative \textit{Kurtosis} values, indicating that the degree of peakedness of the growth and decline of the popularity is flatter than that of the normal distribution. 
Nevertheless, there also exist songs that have extremely large positive values of \textit{Kurtosis}, which gained and lost their popularity very quickly.

\subsection{Effects of debut performance}
\begin{figure}[!t]
\centering
\includegraphics[scale=0.33]{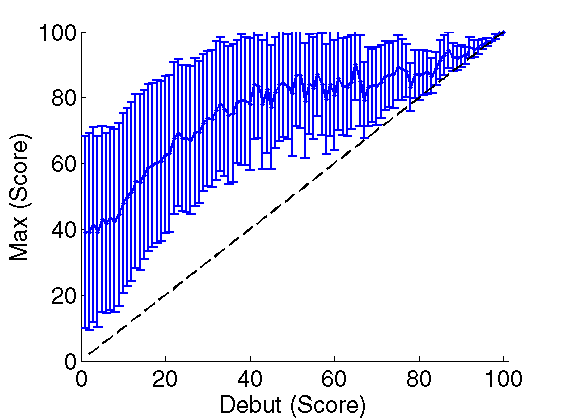}
\caption{Relationship between \textit{Debut} and \textit{Max}. For each value of \textit{Debut}, the average value of \textit{Max} for the songs having the same debut ranking is shown along with the standard deviation as the error bar. The dashed line indicates the cases where \textit{Debut} and \textit{Max} are the same.}
\label{fig:debutmax}
\end{figure}

We now focus on the effects of the initial chart performance of a song, i.e., \textit{Debut}. 
First, we examine the relationship between \textit{Debut} and \textit{Max} in \figurename~\ref{fig:debutmax}. 
It is observed that the \textit{Debut} performance of a song significantly influences its best chart performance. 
It is usually hard to reach high ranks for a song with a low initial rank, but once the initial rank is higher than about 50, then the best performance does not vary much.

\begin{figure}[!t]
\centering
\includegraphics[scale=0.33]{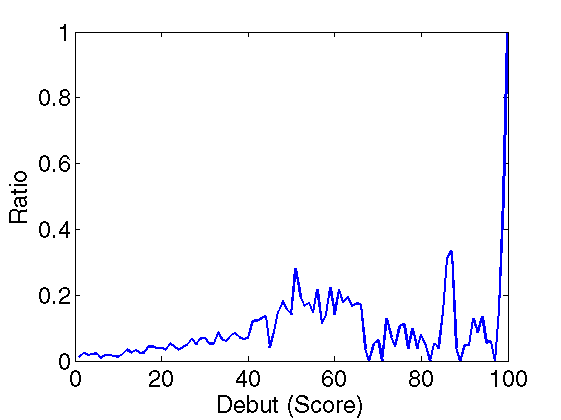}
\caption{Proportion of the number of songs that reached the top rank with respect to \textit{Debut}.}
\label{fig:maxdebut}
\end{figure}
Next, we examine the proportion of the number of songs reaching the highest rank (i.e., \#1) in the chart with respect to \textit{Debut}. 
The result is shown in \figurename~\ref{fig:maxdebut}. 
For the songs whose debut ranks are below a half (i.e., 50), the probability of reaching the top in the chart gradually increases with respect to the debut rank. 
However, once the debut rank is above a half, no significant correlation is observed between the debut rank and the probability of reaching the top. 
Surprisingly, only a half of the songs that debuted in second place in the chart succeeded in reaching the top.

\subsection{Temporal analysis} \label{Temporal}
We examine how the popularity metrics changed over time. 
For this, we set four decadal periods, i.e., 1970s, 1980s, 1990s, and 2000s, and obtain the popularity metrics separately for each period. 
Only the songs that appeared and disappeared in the chart within the same period were considered. 
As a result, 5,074, 4,192, 3,570, and 4,375 songs were included in the four periods, respectively. 

\begin{figure}[!t]
\centering
\subfloat[]{\includegraphics[width=1.6in]{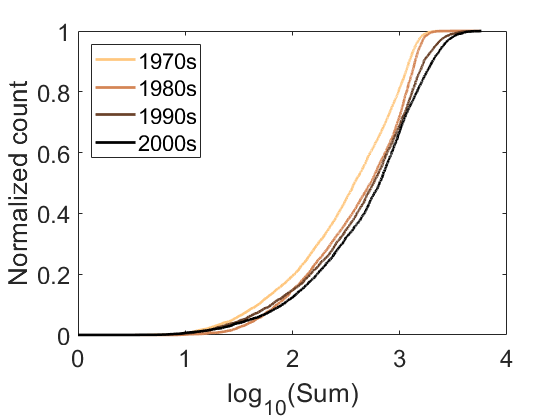}
\label{dec:sum}}
\hfil
\subfloat[]{\includegraphics[width=1.6in]{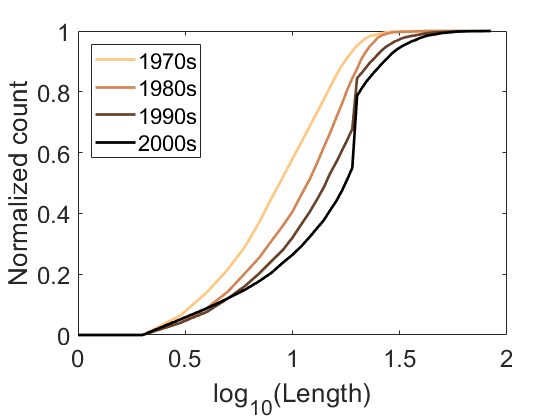}
\label{dec:length}}

\subfloat[]{\includegraphics[width=1.6in]{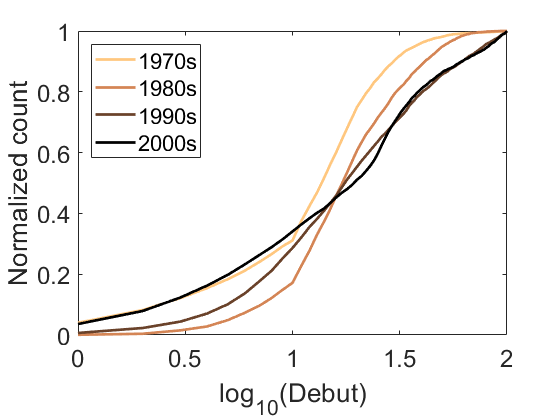}
\label{dec:debut}}
\caption{Cumulative distributions of popularity metrics. (a) \textit{Sum} (b) \textit{Length} (c) \textit{Debut}.}
\label{fig:decade}
\end{figure}

\figurename~\ref{fig:decade} shows the cumulative distributions of \textit{Sum, Length}, and \textit{Debut}. 
Those for the other popularity metrics are omitted because they do not show any noticeable difference among different periods. 
It is observed that a larger portion of songs have larger \textit{Sum} and \textit{Length} values as time goes. 
Larger values of \textit{Length} for more recent decades would be partially responsible for the increase of the proportion of large values of \textit{Sum}.
In \figurename~\ref{dec:length}, abrupt increases at week 20 are observed only for 1990s and 2000s, because the recurrent rule has applied only after 1991. 
In \figurename~\ref{dec:debut}, the distribution of \textit{Debut} becomes more gradual as time goes. 
To sum up, songs in more recent years tend to have appeared in the chart at more various initial ranks and stayed longer in the chart. 

\section{Prediction of Popularity Metrics} \label{Prediction}
We build models that predict the popularity metrics using musical features. 
The complexity features, and two other types of acoustic features used in previous studies are explained. 
Then, the experimental setup and results are presented.

\subsection{Comlexity features}
As described in \cite{u1975, complexity2004}, music complexity is highly involved in determining preference and popularity of a song. 
Therefore, we measure complexity of songs and use them to build learning models.
In particular, we adopt the method presented in \cite{SC}, which measures structural changes of three musical components, namely, harmonic, rhythmic, and timbral components of a song.
The method extracts these components from a song, and then calculates complexity features by measuring temporal changes of the components. 
In addition, we consider loudness as another dimension reflecting music complexity, because loudness of a song and its temporal change are related to emotional arousal of listeners.

\subsubsection{Structural change}
The structural change (SC) \cite{SC} basically measures how fast the time sequence of a component changes over time. 
Three components are considered as follows.
\begin{description}[\IEEEsetlabelwidth{$\alpha\omega\pi\theta\mu\beta$}]
\item[\textit{Chroma}]
\textit{Chroma} describes the instantaneous harmony at a particular moment, which is one of the 12 chords (C, C\#, D, ..., B). 
For each signal segment having a length of 0.25 sec, we use the chord estimation algorithm presented in \cite{chroma2010} to obtain the probability distribution over the 12 chords. 
\item[\textit{Rhythm}]
We employ the fluctuation pattern (FP) \cite{fp2003} to quantify the rhythmic signature at a certain moment in a song. 
A 2.97 second-long Hamming window moving one second at a time is applied to the audio signal. 
The windowed audio segment is further divided into 256 frames (containing 512 samples for 44.1 kHz sampling). 
For each frame, the fast Fourier transform (FFT) is applied and the periodicity histogram \cite{fp2003} of each frequency band is measured to obtain FPs of the frames. 
The mean of the 256 FPs is defined as the rhythm component of the segment.
\item[\textit{Timbre}]
The timbral components of a song are mostly determined by its frequency components. 
Thus, we employ the Mel-frequency cepstral coefficients (MFCCs) \cite{mfcc} that are popular for perceptual representation of spectral aspects of audio signals. 
As in the case of rhythm, we apply a 2.97 second-long Hamming window to the given audio signal and then divide the windowed segment into 256 frames. 
We use 36 Mel-frequency spaced bins for each frame. 
Finally, the MFCCs from the 256 frames are averaged. 
\end{description}

We compute SC on each component to obtain complexity features over time. 
The SC for the $i$th segment with an observation window having a length of \begin{math}w_j=2^{j-1}\end{math} is given by \begin{equation}{SC}_{ij} = JSD(s_{i-w_j:i-1}, s_{i:i+w_j-1})\end{equation} where $JSD(x, y)$ is the Jenson-Shannon divergence between $x$ and $y$ and $s_{a:b}$ is the sum of the values of $s$ for the $a$th to $b$th segments. 
The observation window determines the temporal scope measuring complexity. 
We set \begin{math}j = 3, 4, ..., 8\end{math} for \textit{chroma} and \begin{math}j = 1, 2, ..., 6\end{math} for \textit{rhythm} and \textit{timbre}, which correspond to window lengths of 1 to 32 sec. 
Finally, we take the mean value of SC over time to obtain a single value for each feature and each window size.
As a result, six complexity features for the six observation window lengths are obtained for each of the \textit{Chroma}, \textit{Rhythm}, and \textit{Timbre} components: \textit{Chroma1} to \textit{Chroma6}, \textit{Rhythm1} to \textit{Rhythm6}, and \textit{Timbre1} to \textit{Timbre6}. 

\subsubsection{Arousal}
We use another kind of complexity feature based on the psychoacoustic theory, named \textit{Arousal}. 
It has been known that the loudness of music is correlated to the arousal dimension of emotion \cite{arousal2004}. 
And, emotion is an important factor determining music preference \cite{emotion2010}.
Therefore, we can assume that the musical preference (and consequently music popularity) is influenced by arousal. 

We calculate the short-time magnitude (i.e., sum of the absolute signal values) for a segment obtained using a 2.97-second long Hamming window moving one second at a time. 
We then calculate the mean (\textit{ArousalMean}) and standard deviation (\textit{ArousalStd}) of the short-time magnitudes over time to measure the average and variation of loudness, respectively. 
\subsection{MPEG features}
The MPEG-7 Audio standard\footnote{\url{http://mpeg.chiariglione.org/standards/mpeg-7/audio}} provides tools to measure various spectral and temporal characteristics of audio signals. 
MPEG-7 Audio features are extracted by using the bag-of-frame (BOF) approach, which divides the audio signal into segments, extracts audio features, and summarizes all segments using statistical measures such as sum and standard deviation. 
We use the MPEG-7 Reference Software\footnote{\url{http://mpeg.chiariglione.org/standards/mpeg-7/reference-software}} to obtain 82 MPEG-7 Audio features from each song. 
\subsection{MFCC features}
In \cite{hit2005}, features based on MFCCs were used to measure the spectral characteristics of songs for hit song prediction. 
We also employ them by implementing the feature extraction procedure in \cite{hit2005}.
Each audio signal is divided into segments having a length of 0.025 sec with an overlap of 0.015 sec between adjacent segments, and 20 MFCCs are extracted for each segment. 
For the extracted MFCCs of the songs in the training data set, k-means clustering is conducted, from which 32 cluster centroids are obtained, in order to reduce noise and obtain compact features. 
These centroids represent the most common sound characteristics found in the training data. 
When a song for test is given, the minimum distance centroids for its MFCC vectors are found and the normalized frequencies of the 32 clusters are obtained as the features of the song.

\subsection{Experimental setup} \label{Setup}
\subsubsection{Data}
Listeners' preference to songs tends to change over time (e.g., popular genres in different periods). 
In addition, as shown in Section~\ref{Temporal}, the trends for some popularity metrics have changed. 
Thus, popularity prediction over a long time period such as several decades would not be of much interest. 
Considering this, we conduct our experiments with the songs within the recent five years. 
Specifically, the 1,264 songs appearing in the Billboard Hot 100 chart during 253 weeks between 13 June 2009 and 19 April 2014 are used for our experiments. 
The first 70\% of the period, until 11 November 2012 (including 864 songs), is considered for training classifiers, the first half of the rest (200 songs) for validation, and the last half (200 songs) for testing. 
In other words, we examine whether the prediction of ``future popularity'' is feasible.

\subsubsection{Setup}
We use support vector machines (SVMs) as classifiers\footnote{We present only the results of the most effective classifier in this section, but complete results using other classifiers are shown in Appendix.}.
We train SVMs with the extracted features to perform binary classification of each popularity metric. 
The boundary of the two classes is set to the median value of each popularity metric in the training data set. 
The radial basis function (RBF) is used as the kernel function of the SVMs. 
The values of the penalty factor and the width of the RBF function determined empirically for the validation dataset. 
We use the libsvm package\footnote{\url{https://www.csie.ntu.edu.tw/~cjlin/libsvm/}} \cite{libsvm}. 

We design three different experiments to investigate popularity prediction performance of the features: \begin{enumerate}
\item Prediction using each single complexity feature
\item Prediction using each feature group
\item Prediction using combined features.\end{enumerate}
In the first experiment, the classifier is trained with a single feature to examine whether each complexity feature can predict popularity metrics and which feature is effective for prediction. 
The second experiment compares the prediction performance of each feature group, i.e., \textit{Complexity, MPEG}, and \textit{MFCC}. 
Finally, the third experiment examines performance improvement by additional use of \textit{MFCC} and/or \textit{MPEG} together with \textit{Complexity}. 

In order to compensate for different numbers of data in the two classes, the classification performance is measured in terms of the balanced accuracy, which is defined as \begin{equation}BA = {1\over 2}({tp\over tp+fn}+{tn\over tn+fp}) \end{equation} where $tp$, $tn$, $fp$, and $fn$ are the numbers of true positives (hit), true negatives (correct rejection), false positives (false alarm), and false negatives (miss), respectively. 

\subsection{Results}
\subsubsection{Single feature-based prediction}
\begin{figure*}[!t]
\centering
\includegraphics[scale=0.33]{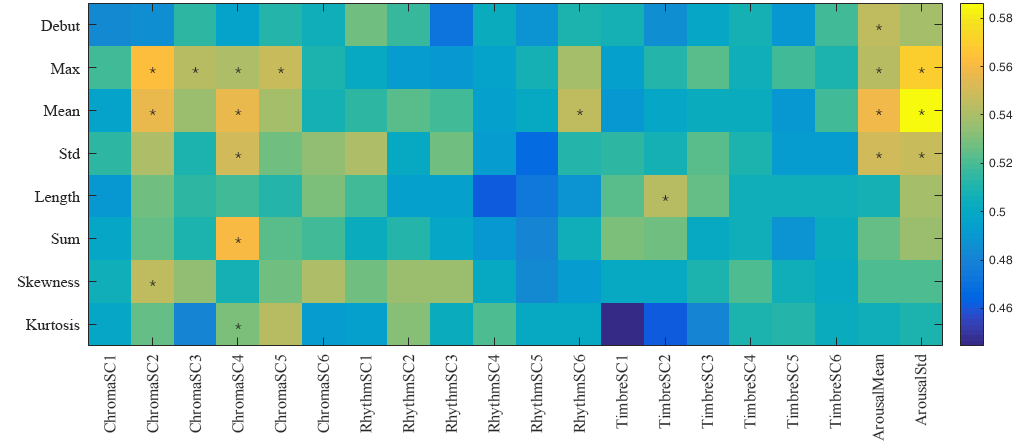}
\caption{Results of popularity prediction using single \textit{Complexity} features in terms of balanced accuracy. The cases where the accuracy is statistically significantly different from random chance are marked with `*'.}
\label{fig:1to1}
\end{figure*}
\figurename~\ref{fig:1to1} summarizes the prediction results using single \textit{Complexity} features. 
One-sided t-tests are conducted to examine if the balanced accuracies are significantly different from the random chance (i.e., 0.5). 
The cases showing statistical significance are marked with `*' in the figure. 
In total, 11.9\% of all cases are shown to be statistically significant, which indicates that some complexity features are effective for predicting popularity metrics even when used singly. 

Among the four components of complexity, features from \textit{Chroma }(\textit{ChromaSC1 }to \textit{ChromaSC6}) are the most effective; 20.8\% of the cases using the chroma-based features show statistically significant performance. 
The \textit{Rhythm }and \textit{Timbre }features are not as effective as the \textit{Chroma }features. 
Interestingly, the \textit{Arousal }features, which are based on simple loudness measurement, perform well, showing significant performance for \textit{Debut}, \textit{Max}, \textit{Mean}, and \textit{Std}. 

\textit{Chroma} is related to the flow of a song such as melody, mood, and chord, and listeners prefer appropriate change of such flow. 
Although \textit{Rhythm} and \textit{Timbre} features are less effective, they are complementary with each other; the features from \textit{Timbre} are effective for predicting \textit{Length}, while those from \textit{Rhythm} are effective for \textit{Mean}. 

In the viewpoint of popularity metrics, \textit{Debut} and \textit{Kurtosis} are difficult to predict with single \textit{Complexity} features. 
Only \textit{Arousal} features are effective for \textit{Debut}. 

When the performance with respect to the window size is examined, there exists a slight tendency that smaller windows are more effective for \textit{Rhythm} and \textit{Timbre} than larger ones, whereas the window size does not affect the performance of \textit{Chroma} much as long as it is not the smallest or largest. 
It is thought that rapid changes of rhythm or timbre are easily captured by listeners, thus have impact on the listeners' preference. 
However, chromatic changes due to chord alteration over both small and large time scales are expected in most songs, thus a wide range of window sizes appear effective.

\subsubsection{Feature group-based prediction}
\begin{figure}[!t]
\centering
\includegraphics[scale=0.27]{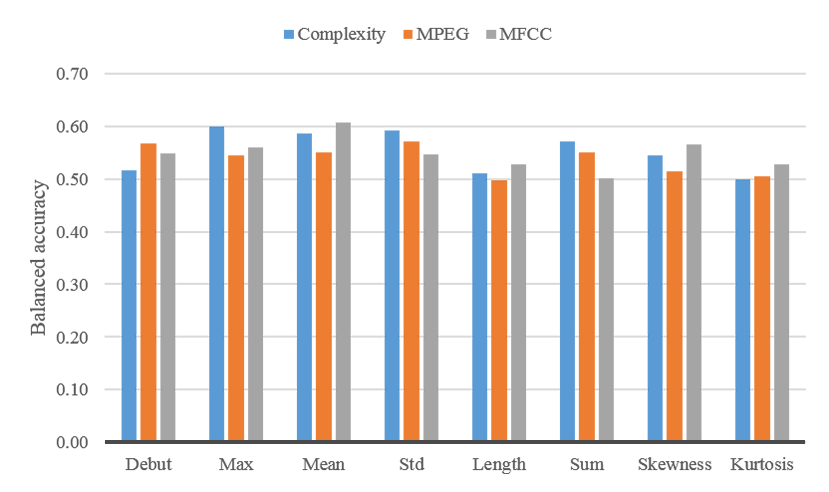}
\caption{Results of prediction using each feature group.}
\label{fig:group}
\end{figure}

The prediction results of each feature group are shown in \figurename~\ref{fig:group}. 
For \textit{Debut}, \textit{Length}, and \textit{Kurtosis}, the prediction performance of each feature group is overall inferior to that for the other popularity metrics. 
In fact, the debut ranking of a song may be more influenced by awareness and popularity of the artist, promotion, or social trend, as in other cultural products \cite{market2006, boxoffice1996, boxoffice2010}. 

The group of \textit{Complexity} is the most effective among the three groups in predicting \textit{Max}, \textit{Std}, and \textit{Sum}, for which relatively good performance was observed in \figurename~\ref{fig:1to1}. 
\textit{MFCC} shows the best prediction performance for \textit{Length, Mean, Skewness}, and \textit{Kurtosis}, although the accuracies are not high for \textit{Length} and \textit{Kurtosis}. 
\textit{MPEG} ranks first only for \textit{Debut}. 

These results demonstrate that the \textit{Complexity} features capture popularity-related characteristics of songs more effectively than the \textit{MFCC} and \textit{MPEG} features that are designed to extract rather general acoustic characteristics.
\textit{Complexity} and \textit{MFCC} showing better performance than \textit{MPEG}, perceptual features are more suitable for popularity prediction.

\subsubsection{Prediction using combined features}
\begin{figure}[!t]
\centering
\includegraphics[scale=0.27]{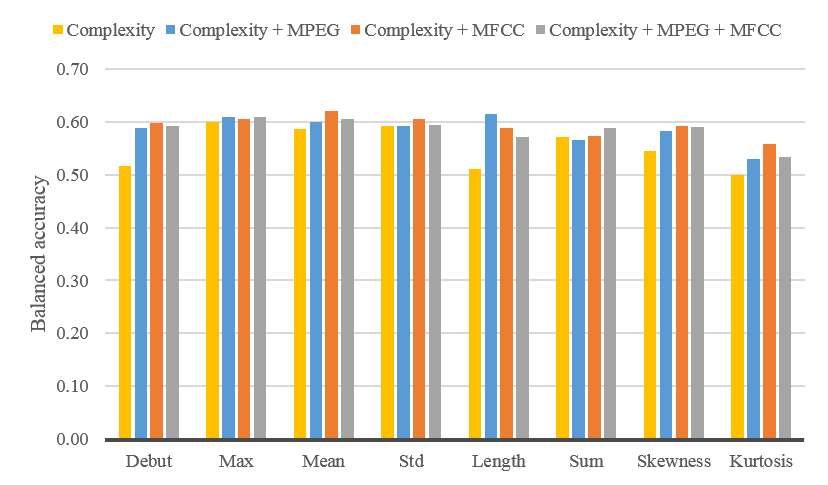}
\caption{Results of prediction using combined features.}
\label{fig:fusion}
\end{figure}

\begin{table}[t]
\centering
\caption{Confusion matrix showing the performance of \textit{Complexity} and \textit{MFCC}}
\label{tab:confusion}
\begin{tabular}{|cc|C{1.1cm}C{1.1cm}|}
\firsthline
& & \multicolumn{2}{c|}{\textit{Complexity}} \\
%\cline{3-4}
& & hit & miss \\
\hline
\multirow{2}*{{\textit{MFCC}}} & hit & 0.391 & 0.176 \\
%\multirow{2}*{\rotatebox[origin=c]{90}{\textit{MFCC}}} & hit & 625 & 281\\
%\cline{2-4}
& miss & 0.176 & 0.258 \\
\lasthline
\end{tabular}
\end{table}
Table~\ref{tab:confusion} shows the confusion matrix examining how \textit{Complexity} and \textit{MFCC} behave in feature group-based prediction. 
The cases where one group produces correct classification results but the other does not take about 35\% (the off-diagonal elements in the matrix), for which we can expect complementarity when they are combined.

\figurename~\ref{fig:fusion} summarizes the balanced accuracies of prediction when the \textit{MPEG} and \textit{MFCC} features are combined with the \textit{Complexity} features. 
In most cases, additional use of \textit{MPEG} and/or \textit{MFCC} improves the performance. 
In particular, the improvement is prominent for \textit{Debut}, \textit{Length}, and \textit{Kurtosis}. 
Overall, the combination of \textit{Complexity} and \textit{MFCC} appears the best, recording an average accuracy of 59.2\%. 
Further combination of \textit{MPEG} deteriorates the performance in most cases, which seems to be due to the excessively increased feature dimensionality.

The accuracies in \figurename~\ref{fig:fusion} may look quite low. 
But it is thought that the problem of popularity prediction is very challenging, as also shown in previous studies. 
For instance, in \cite{hit2005}, a problem of classifying whether a song will rank first was considered, for which relatively low performance was reported (0.66 in terms of area under curve (ROC)). 
In \cite{hit2012}, no statistically significant performance in comparison to random chance was obtained for a three-class popularity classification problem. 
In our case, on the other hand, one-sided t-tests reveal that the accuracies for \textit{Complexity}+\textit{MFCC} are statistically significant except for \textit{Kurtosis}. 

\subsubsection{Incorporating \textit{Debut} as a feature}
\begin{figure}[!t]
\centering
\includegraphics[scale=0.27]{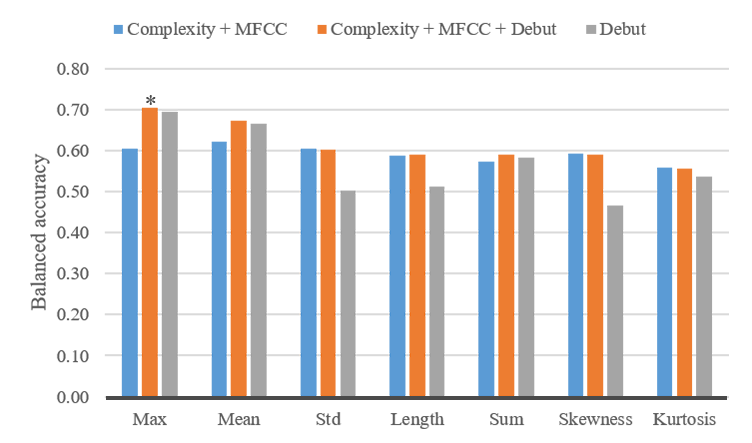}
\caption{Prediction results when \textit{Debut} is used as an input feature for prediction. For \textit{Max}, the improvement is statistically significant ($p=0.046)$.}
\label{fig:efusion}
\end{figure}
Finally, we examine a different scenario, where popularity metrics of a song are predicted after it has debuted in the chart. 
While the acoustic features-based popularity prediction shown above are useful even before a song is released to the public, it is also interesting how the popularity of a song will evolve once we know its debut performance in the chart.

For this, the debut score of a song is used as a feature in addition to the acoustic features for prediction. 
Since the combination of \textit{Complexity} and \textit{MFCC} showed the best performance in \figurename~\ref{fig:fusion}, we evaluate the performance of the combination of \textit{Complexity}, \textit{MFCC}, and \textit{Debut}, which is shown in \figurename~\ref{fig:efusion}. 
A significant increase of the accuracy is observed for \textit{Max} (absolute improvement by 9.9\%) due to the additional use of the debut score. 
The strong correlation between \textit{Debut} and \textit{Max}, which was shown in \figurename~\ref{fig:debutmax}, and thereby a high accuracy using even only \textit{Debut} for \textit{Max}, seems to contribute to the performance improvement.
We also observe slight performance improvement for \textit{Mean}. 
For the rest, there is almost no performance difference.

\section{Conclusions} \label{Conclusions}
In this paper, we have defined popularity metrics, analyzed their properties from real chart data, and automatically predicted them using acoustic features. 
To consider various aspects of popularity, we defined eight popularity metrics: \textit{Debut}, \textit{Max}, \textit{Mean}, \textit{Std}, \textit{Length}, \textit{Sum}, \textit{Skewness}, and \textit{Kurtosis} from rankings of a song in the chart, which are complementary with each other.

We examined the characteristics of the popularity metrics with the real chart data for 16,686 songs ranked in the Billboard Hot 100 chart between 1970 and 2014.
Each popularity metric showed a distinct distribution, from which noteworthy observations were made. 
Most songs scored low ranks at the beginning in the chart. 
Although many songs reached the highest rank at least once, their average scores were distributed in the middle range of the chart.
The growth pattern of popularity was usually gradual, rising slowly and falling quickly.
The maximum rank of a song was highly related to its debut performance. 
We also investigated the popularity metrics in different decades.
It was observed that in more recent years, more songs debuted in the chart and stayed longer. 

Prediction of the popularity metrics was conducted using acoustic features.
We proposed to use \textit{Complexity} features including musical and arousal components for popularity prediction based on the theory describing close relationship between complexity and preference. 
We performed binary classification using SVMs with 1,264 songs in the Billboard Hot 100 chart for 253 weeks.
Single features from \textit{Chroma} and \textit{Arousal} performed well, which indicates that these components are related to preference of listeners. 
Overall, \textit{Complexity} was superior to \textit{MFCC} and \textit{MPEG}. 
Combining \textit{Complexity} and \textit{MFCC} improved prediction performance for most popularity metrics. 
It was also shown that the information of the debut score is effective to increase further the accuracy for \textit{Max}. 

To sum up, we have investigated what popularity of a song is, how the popularity metrics look in the real world, and whether they are predictable using acoustic features.
In the future, it would be necessary to attempt to improve the prediction performance, e.g., using deep neural networks.

\appendix \label{Appendix}
\begin{table*}[t]
	\centering
	\caption{Balanced accuracy of different classifiers}
	\label{tab:acc}
	\begin{tabular}{cl|C{1.1cm}C{1.1cm}C{1.1cm}C{1.1cm}C{1.1cm}C{1.1cm}C{1.1cm}C{1.1cm}}
		\firsthline
		Classifier & Feature	& Debut & Max & Mean & Std & Length & Sum & Skewness & Kurtosis \\
		\hline
		\multirow{3}*{SVM}	& Complexity &	0.52&	0.60&	0.59&	0.59&	0.51&	0.57&	0.54&	0.50\\
		&	MPEG&	0.57&	0.54&	0.55&	0.57&	0.50&	0.55&	0.51&	0.51\\
		&	MFCC&	0.55&	0.56&	0.61&	0.55&	0.53&	0.50&	0.57&	0.53\\
		
		\hline
		
		\multirow{3}*{LR}	& Complexity&0.51	&0.59	&0.60	&0.57	&0.57	&0.56	&0.52	&0.50\\
		& MPEG	&0.47	&0.49	&0.50	&0.48	&0.55	&0.54	&0.47	&0.47\\
		& MFCC	&0.53	&0.56	&0.59	&0.53	&0.54	&0.56	&0.61	&0.48\\
		
		\hline
		\multirow{3}*{DT}	& Complexity&0.52&	0.51&	0.60&	0.48&	0.47&	0.47&	0.47&	0.41 \\
		& MPEG&0.53&	0.56&	0.55&	0.49&	0.48&	0.55&	0.46&	0.56 \\
		& MFCC&0.53&	0.49&	0.54&	0.51&	0.50&	0.53&	0.49&	0.50 \\
		\hline
		\multirow{3}*{NN}	& Complexity&0.45&	0.51&	0.56&	0.50&	0.55&	0.52&	0.57&	0.49\\
		& MPEG&0.49&	0.53&	0.54&	0.57&	0.53&	0.50&	0.48&	0.52\\
		& MFCC&0.56&	0.59&	0.55&	0.53&	0.47&	0.50&	0.54&	0.53\\
		\hline
		CNN & & 0.52 & 0.52 & 0.53 & 0.48 & 0.49 & 0.55 & 0.48 & 0.53 \\
		\lasthline
	\end{tabular}
\end{table*}
We present the complete results of feature group-based popularity metric prediction using various classifiers. 
The employed classifiers are decision tree (DT), logistic regression (LR), neural network with a single hidden layer having sigmoidal neurons (NN), and CNN. 
The number of hidden neurons in NN is determined empirically for the validation dataset.
NN is trained with the Levenberg-Marquardt algorithm that is one of the fastest learning algorithms. 
For the CNN, the model presented in \cite{yang2017CNN} was adopted, where a fully connected layer is used for the final layer. 
The input of the CNN is the 128-band Mel-spectrogram of the 120-second-long middle part of each audio signal.

The performance in terms of balanced accuracy of each classifier is shown in Table~\ref{tab:acc}.
On average, SVM shows the best performance among the classifiers with consistently good accuracies for the three types of features when all the popularity metrics are considered, although some classifiers show better performance for some cases (e.g., LR using \textit{MFCC} for \textit{Skewness}). 
The deep learning approach performs poor, probably due to the insufficient amount of data for training.

%\section*{Acknowledgment}
%This research was supported by the MISP (Ministry of Science, ICT and Future Planning), Korea, under the ``IT Consilience Creative Program" (IITP-2015-R0346-15-1008) supervised by the IITP (Institute for Information \& Communications Technology Promotion).

% Can use something like this to put references on a page
% by themselves when using endfloat and the captionsoff option.
\ifCLASSOPTIONcaptionsoff
  \newpage
\fi

% trigger a \newpage just before the given reference
% number - used to balance the columns on the last page
% adjust value as needed - may need to be readjusted if
% the document is modified later
%\IEEEtriggeratref{8}
% The "triggered" command can be changed if desired: 
%\IEEEtriggercmd{\enlargethispage{-5in}}

% references section

% can use a bibliography generated by BibTeX as a .bbl file
% BibTeX documentation can be easily obtained at:
% http://mirror.ctan.org/biblio/bibtex/contrib/doc/
% The IEEEtran BibTeX style support page is at:
% http://www.michaelshell.org/tex/ieeetran/bibtex/
\bibliographystyle{IEEEtran}
% argument is your BibTeX string definitions and bibliography database(s)
\bibliography{bare_jrnl}
%
% <OR> manually copy in the resultant .bbl file
% set second argument of \begin to the number of references
% (used to reserve space for the reference number labels box)
%\begin{thebibliography}{1}
%
%\bibitem{IEEEhowto:kopka}
%H.~Kopka and P.~W. Daly, \emph{A Guide to \LaTeX}, 3rd~ed.\hskip 1em plus
%  0.5em minus 0.4em\relax Harlow, England: Addison-Wesley, 1999.
%
%\end{thebibliography}

% biography section
% 
% If you have an EPS/PDF photo (graphicx package needed) extra braces are
% needed around the contents of the optional argument to biography to prevent
% the LaTeX parser from getting confused when it sees the complicated
% \includegraphics command within an optional argument. (You could create
% your own custom macro containing the \includegraphics command to make things
% simpler here.)
%\begin{IEEEbiography}[{\includegraphics[width=1in,height=1.25in,clip,keepaspectratio]{mshell}}]{Michael Shell}
% or if you just want to reserve a space for a photo:

\begin{IEEEbiography}[{\includegraphics[width=1in,height=1.25in,clip,keepaspectratio]{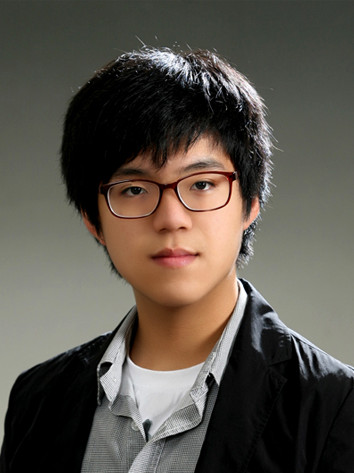}}]{Junghyuk Lee}
received his B.S. degree from the School of Integrated Technology at Yonsei Uni- versity, Korea, in 2015, where he is currently working toward the Ph.D. degree. His research interests include multimedia signal processing and wide color gamut imaging.
\end{IEEEbiography}

\begin{IEEEbiography}[{\includegraphics[width=1in,height=1.25in,clip,keepaspectratio]{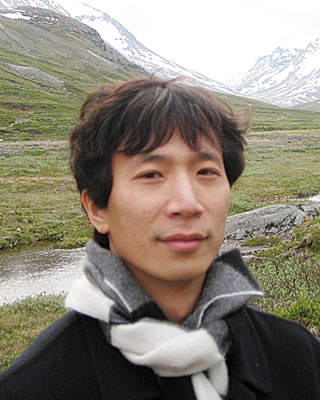}}]{Jong-Seok Lee}
(M06-SM14) received his Ph.D. degree in electrical engineering and computer science in 2006 from KAIST, Korea. From 2008 to 2011, he worked as a research scientist at Swiss Federal Institute of Technology in Lausanne (EPFL), Switzerland. Currently, he is an associate professor in the School of Integrated Technology at Yonsei University, Korea. His research interests include multimedia signal processing and machine learning. He is an author or co-author of over 100 publications. He serves as an editor for the IEEE Communications Magazine and Signal Processing: Image Communication.

\end{IEEEbiography}
%% if you will not have a photo at all:
%\begin{IEEEbiographynophoto}{John Doe}
%Biography text here.
%\end{IEEEbiographynophoto}
%
%% insert where needed to balance the two columns on the last page with
%% biographies
%%\newpage
%
%\begin{IEEEbiographynophoto}{Jane Doe}
%Biography text here.
%\end{IEEEbiographynophoto}

% You can push biographies down or up by placing
% a \vfill before or after them. The appropriate
% use of \vfill depends on what kind of text is
% on the last page and whether or not the columns
% are being equalized.

%\vfill

% Can be used to pull up biographies so that the bottom of the last one
% is flush with the other column.
%\enlargethispage{-5in}

% that's all folks
\end{document}